\documentclass[11pt,a4paper]{article}
\pdfoutput=1
\usepackage{jheppub}

\RequirePackage[utf8]{inputenc}

\usepackage{amsfonts}
\usepackage{amssymb}
\usepackage{comment}
\usepackage{graphicx}
\usepackage{amsmath}
\usepackage{tabu}
\usepackage{dsfont}
\usepackage{float}
\usepackage{amsthm}
\usepackage{slashed}
\usepackage{setspace}
\usepackage[labelformat=simple]{subcaption}

\usepackage{pgfplots}
\usepackage{tikz}
\usepackage{tikz-cd}
\usetikzlibrary{shapes.misc,shadows,fit,decorations.pathmorphing,positioning,trees,decorations.markings,decorations.pathreplacing,calc,shapes,patterns,arrows,positioning,arrows.meta}
\usepackage{xcolor}
\usepackage{pgfplots}
\usepackage{pgfplotstable}
\usepackage{xcolor}
\usepackage[boxsize=0.5em,aligntableaux=center]{ytableau}
\usepackage{makecell}
\usepackage{diagbox}
\usepackage{environ}
\usepackage[utf8]{inputenc}
\usepackage{amsmath, amsthm, amssymb, amsfonts}
\usepackage{amsmath, amsthm, amssymb, amsfonts}
\usepackage{multirow}
\usepackage{graphicx}
\usepackage{float}
\usepackage[numbers,sort&compress]{natbib}
\usepackage{jheppub}
\usepackage{enumerate}
\usepackage{enumitem}
\usepackage[capitalize,sort]{cleveref}
\usepackage{hhline}
\usepackage{mathtools}
\usepackage{longtable}
\usepackage{makecell}
\usepackage{tabularx}
\usepackage{cellspace}
\usepackage{alphalph}
\usepackage{lscape}

\crefname{figure}{Figure}{Figures}
\crefname{table}{Table}{Tables}

\def\ov{\overline}

\DeclareMathOperator{\sign}{sign}

\DeclareMathOperator{\GCD}{GCD}
\newcommand{\PP}{\mathbb{P}}

\newcommand{\coma}{\, , \quad}
\newcommand{\fstop}{\, .}

\renewcommand{\epsilon}{\varepsilon}

\makeatletter
\newsavebox{\measure@tikzpicture}
\NewEnviron{scaletikzpicturetowidth}[1]{%
  \def\tikz@width{#1}%
  \begin{lrbox}{\measure@tikzpicture}%
  \BODY
  \end{lrbox}%
  \pgfmathparse{#1/\wd\measure@tikzpicture}%
  \BODY
}
\makeatother

\definecolor{col1}{HTML}{f94144}
\definecolor{col2}{HTML}{f3722c}
\definecolor{col3}{HTML}{f8961e}
\definecolor{col4}{HTML}{f9844a}
\definecolor{col5}{HTML}{f9c74f}
\definecolor{col6}{HTML}{90be6d}
\definecolor{col7}{HTML}{43aa8b}
\definecolor{col8}{HTML}{4d908e}
\definecolor{col9}{HTML}{577590}
\definecolor{col10}{HTML}{277da1}

\catcode`\@=11   
\newdimen\@rotdimen
\newbox\@rotbox  

\def\@vspec#1{\special{ps:#1}}
\def\@rotstart#1{\@vspec{gsave currentpoint currentpoint translate
		#1 neg exch neg exch translate}}
\def\@rotfinish{\@vspec{currentpoint grestore moveto}}
\def\@rotr#1{\@rotdimen=\ht#1\advance\@rotdimen by\dp#1%
	\hbox to\@rotdimen{\hskip\ht#1\vbox to\wd#1{\@rotstart{90 rotate}%
			\box#1\vss}\hss}\@rotfinish}
\def\@rotl#1{\@rotdimen=\ht#1\advance\@rotdimen by\dp#1%
	\hbox to\@rotdimen{\vbox to\wd#1{\vskip\wd#1\@rotstart{270 rotate}%
			\box#1\vss}\hss}\@rotfinish}%
\def\@rotu#1{\@rotdimen=\ht#1\advance\@rotdimen by\dp#1%
	\hbox to\wd#1{\hskip\wd#1\vbox to\@rotdimen{\vskip\@rotdimen
			\@rotstart{-1 dup scale}\box#1\vss}\hss}\@rotfinish}%
\def\@rotf#1{\hbox to\wd#1{\hskip\wd#1\@rotstart{-1 1 scale}%
		\box#1\hss}\@rotfinish}%
\def\rotate{\@ifnextchar[{\@rotate}{\@rotate[l]}}
\def\@rotate[#1]#2{\setbox\@rotbox=\hbox{#2}\@nameuse{@rot#1}\@rotbox}

\catcode`\@=12

\tikzset{
    partial ellipse/.style args={#1:#2:#3}{
        insert path={+ (#1:#3) arc (#1:#2:#3)}
    }
}

\allowdisplaybreaks[2]
\numberwithin{equation}{section}

\usepackage[framemethod=TikZ]{mdframed}
\mdfsetup{skipabove=\topskip, skipbelow=\topskip}

\newcounter{equ}[section]

\newcounter{Boxequ}[section]

\makeatletter
\def\user@resume{resume}
\def\user@intermezzo{intermezzo}
\newcounter{previousequation}
\newcounter{lastsubequation}
\newcounter{savedparentequation}
\makeatother

\preprint{ZMP-HH/22-4}

\title{Systematics of perturbatively flat flux vacua for CICYs}

\author[a]{Federico Carta,}
\author[b]{Alessandro Mininno,}
\author[c]{Pramod Shukla}

\affiliation[a]{Department of Mathematical Sciences, Durham University, \\ Durham, DH1 3LE, United Kingdom}
\affiliation[b]{II. Institut f\"ur Theoretische Physik, Universit\"at Hamburg,\\
Luruper Chaussee 149, 22607 Hamburg, Germany}
\affiliation[c]{\small ICTP, Strada Costiera 11, Trieste 34151, Italy}

\emailAdd{federico.carta@durham.ac.uk}
\emailAdd{alessandro.mininno@desy.de}
\emailAdd{pramodmaths@gmail.com}

\abstract{In this paper, we extend the analysis of scanning the perturbatively flat flux vacua (PFFV) for the type IIB orientifold compactifications on the mirror of the projective complete intersection Calabi-Yau (pCICY) 3-folds, which are realized as hypersurfaces in the product of complex projective spaces. The main objective of this scan is to investigate the behaviour of PFFV depending on the nature of CY 3-folds in the light of the observations made in \cite{Carta:2021kpk} where it has been found that K3-fibered CY 3-folds have significantly large number of physical vacua as compared to other geometries. For this purpose, we present the PFFV statistics for all the 36 pCICYs with $h^{1,1}=2$ and classify them into two categories of being K3-fibered model and non K3-fibered model. We subsequently confirm that all the K3-fibered models have a significantly large number of PFFV leading to physical vacua by fixing the axio-dilaton by non-perturbative effects, while only a couple of non K3-fibered models have such physical vacua. For $h^{1,1}=2$ case, we have found that there are five pCICY $3$-folds with the suitable exchange symmetry leading to the so-called exponentially flat flux vacua (EFFV) which are protected against non-perturbative prepotential effects as well. By exploring the underlying exchange symmetries in the favorable CY 3-folds with $h^{1,1} \geq 3$ in the dataset of 7820 pCICYs, we have found that there are only 13 spaces which can result in EFFV configurations, and therefore most of the CY 3-folds are a priory suitable for fixing the dilaton valley of the flat vacua using the non-perturbative prepotential contributions.
}


\begin{document}

\maketitle

\bigskip

\section{Introduction}
\label{sec_intro}

Since critical superstring theory is defined in 10 spacetime dimensions, one needs to compactify the theory on a 6-manifold in order to extract the relevant low energy effective field theory in 4 dimensions. In this context, one of the main open problems regards the stabilization of moduli, scalar fields present in the effective theory, whose vacuum expectation value (VEV) is related to shape and size of the compactification space.

In this paper, we focus on IIB string theory compactified on a Calabi-Yau (CY) orientifold $X$. In such setup, the moduli stabilization typically is performed in two steps. Complex structure moduli and the axio-dilaton are stabilized by background fluxes at higher energy \cite{Gukov:1999ya,Giddings:2001yu}. Due to the no-scale property of the prepotential \cite{Cremmer:1983bf}, this leaves all directions corresponding to K\"ahler moduli flat. Those K\"ahler moduli are then stabilized at much lower energies by the introduction of subleading effects, such as higher string-loop corrections, or non-perturbative effects as the presence of Euclidean 3-branes, or the worldvolume dynamic of a D7-brane wrapping a rigid divisor and undergoing gaugino condensation \cite{Witten:1996bn,Balasubramanian:2005zx}. Famous K\"ahler moduli stabilization approaches include KKLT \cite{Kachru:2003aw}, LVS \cite{Balasubramanian:2005zx}, racetrack \cite{Denef:2004dm}, K\"ahler uplift \cite{Westphal:2006tn}, perturbative LVS \cite{Antoniadis:2018hqy,Antoniadis:2018ngr,Antoniadis:2019rkh} and others.

Let $F_3$ (resp. $H_3$) be the Ramond-Ramond (resp. Neveu-Schwarz) three-form flux, $S$ be the axio-dilaton, and $\Omega_3$ the nowhere vanishing holomorphic three-form which $X$ admits by definition, being Calabi-Yau. After complex structure moduli and axio-dilaton stabilization by the so-called Gukov-Vafa-Witten (GVW) superpotential \cite{Gukov:1999ya}, one is left with a constant term, usually called as flux superpotential, and given by
\begin{equation}
W_0=\sqrt{\frac{2}{\pi}}\left\langle e^{K/2}\int_X\left(F_3-SH_3\right)\wedge \Omega_3\right\rangle\fstop
\end{equation}
The viability of the subsequent K\"ahler moduli stabilization approach depends, sometimes, on the explicit value of $W_0$. In particular, in the KKLT context it is necessary, for control reasons, to have a relatively small flux superpotential. On these lines, there have been some significant amount of attraction on the magnitude of $W_0$ for model builders \cite{Cicoli:2013swa,Louis:2012nb,Demirtas:2019sip,Demirtas:2020ffz, Broeckel:2021uty, Bastian:2021hpc, Alvarez-Garcia:2020pxd, Grimm:2021ckh,Carta:2021kpk}.
In \cite{Demirtas:2019sip} an explicit algorithm to achieve small superpotentials was put forward. Schematically, first one looks for a flux choice that will generate a scalar potential with valleys of perturbatively flat supersymmetric vacua, which are in turn lifted by instanton effects. In this context, a systematic scan of such flat vacua leading to physical solutions for the CY with $h^{1,1} = 2$ from the Kreuzer-Skarke database has been presented in \cite{Carta:2021kpk}.

\subsubsection*{Motivation and goals}

Despite being part of the so-called hidden sector, the relics of the compactifying CY 3-folds make them omnipresent in the four-dimensional effective supergravity theory, or any phenomenological model built within that framework. In this context, there are two main CY dataset, namely the Kreuzer-Skarke (KS) database \cite{Kreuzer:2000xy} and the so-called projective complete intersection CY (pCICY) database \cite{Green:1986ck,Candelas:1987kf,Green:1987cr,Gray:2014fla,Anderson:2017aux}. Calabi-Yau manifolds of the pCICY type are realized as the zero-locus of a collection of line bundle sections in an ambient space $\mathcal{A}$ given by a product of complex projective spaces. This database was famously systematized in \cite{Candelas:1987kf}, improved in \cite{Anderson:2017aux}, and contains 7890 different explicit CYs. Out of them, at least 6651 are shown to be topologically distinct \cite{Carta:2021sms}. Furthermore, any other pCICYs which do not belong to the database can be shown to be topologically equivalent to one present in the database, which is therefore complete. Together with the Kreuzer-Skarke (KS) database of hypersurfaces in toric Fano $4$-folds \cite{Kreuzer:2000xy}, and the generalized CICY constructions \cite{Anderson:2015iia}, the original pCICY database is up-to-date one of the three largest and explicitly mapped CY constructions.

The broad motivation and particular goals for the current work can be listed along the following lines:
\begin{itemize}

\item
Despite their extensive use in constructing MSSM-like models, the CY 3-folds of the pCICY dataset have been rarely utilized for moduli stabilization and any subsequent phenomenological purposes such as realizing de-Sitter vacua and inflationary aspects.\footnote{There have been some initiatives along these directions, e.g. for Heterotic moduli stabilization, see \cite{Anderson:2010mh,Anderson:2011cza}, and for moduli stabilization in type IIB setups \cite{Bobkov:2010rf,Carta:2021sms,Carta:2021uwv, Carta:2022web}. In addition, a complete list of explicit classification of CICY orientifolds with non-trivial $(1,1)$-cohomology has been also presented in \cite{Carta:2020ohw}.} Therefore, it makes to us much sense to investigate which CYs in this class can host (flat) flux vacua with a small flux superpotential which being the central ingredients of the KKLT-like moduli stabilization scheme can indeed open a new window for phenomenological applications of pCICYs in this direction. After computing all the divisor topologies on the pCICYs along with investigation of non-perturbative superpotential contributions in a companion work we have already initiated some progress in the direction of K\"ahler moduli stabilization \cite{Carta:2022web}. 

\item Using the CY 3-folds from the KS database, it has been observed in \cite{Carta:2021kpk} that K3-fibered mirror CY 3-folds result in significantly large number of perturbatively flat flux vacua (PFFV). Therefore, it is interesting to investigate this observation for the CY 3-folds of the pCICY database in order to check whether the previous observation is accidental or there is some kind of generality in it, leading to some more deep reasons behind this observation. This observation gets more crucial in the light of the fact that the percentage of K3-fibered CY 3-folds increases very significantly with increasing $h^{1,1}$ not only for the KS database but also for the pCICY database, and therefore it implies that there can be much more CY 3-folds which could possibly be suitable for KKLT-like moduli stabilization.

In this regard, the central aim of this paper is to extensively run the algorithm of \cite{Demirtas:2019sip} by taking  compactification manifold $X$ as the mirror Calabi-Yaus of those belonging to the projective complete intersection CYs (pCICYs) dataset.

\item
However, we consider only a subset of CY 3-folds with $h^{1,1}=2$ from the full pCICY dataset for the purpose of studying this limited aspect. This approach is not only pragmatic for making the overall task simpler but also sufficient to see the pattern that K3-fibered pCICYs are singled out (with a large number of PFFVs) from the full list of pCICYs with $h^{1,1}=2$.

To be specific, we mainly focus on the 36 favorable pCICYs with $h^{1,1}=2$ in our current analysis. We find that only 18 of those give a total of 562 physical vacua via fixing the flat valley through non-perturbative effects, with the lowest value of $|W_0|$ turns out to be $|W_0| \sim 10^{-27}$. Among these $18$ pCICYs, 16 of them are K3-fibered and $2$ of them are not. 

In our detailed analysis using CY 3-folds from the pCICY dataset, we have indeed confirmed that the number of PFFV is dominated by the K3-fibered pCICYs. Since many pCICYs are K3-fibered \cite{Anderson:2017aux}, one expects to have a significant landscape of the PFFV from the pCICY dataset which has a total of 7820 favorable CY $3$-folds.

\item
After performing a detailed scan of PFFV for all the 36 PCICYs with $h^{1,1} = 2$, we explore the possibilities of the so-called exponentially flat flux vacua (EFFV). Such vacua remain flat also considering non-perturbative effects. In \cite{Carta:2021kpk} we have observed that such EFFV arise in the presence of an underlying symmetry in the CY $3$-fold which can be correlated with the so-called non-trivially identical divisors (NIDs) necessary to have an exchange symmetry \cite{Gao:2013pra}. Because of the fact that the possibilities of finding NIDs is quite large in pCICYs it raises a concern of most of the CY 3-folds to result in EFFVs and hence unable to fix the dilaton valley by non-perturbative prepotential effects at all. If it is really the case, then it certainly reduces the suitable CY spaces for the purpose of realizing low $|W_0|$ or KKLT-like mechanism to work. For this reason, it is important to investigate the status of finding EFFVs by considering CY 3-folds with $h^{1,1} \geq 3$.

In this regard, we use the information about the divisor topologies of the pCICYs which has been recently reported in \cite{Carta:2022web}. This enables us to explore the CYs with NIDs which could lead to EFFV configurations, beyond $h^{1,1} =2$ as well. We find five pCICYs with $h^{1,1}=2$ that have this property, for a total of 9543 EFFV configurations for the five CYs, out of a total number of 64520 PFFV configurations arising from all the 36 pCICYs. Subsequently, we also discuss EFFV configurations for pCICY geometries with larger $h^{1,1}$, by investigating the underlying exchange symmetry. In particular, we find that only those pCICYs that have a single topology for all their coordinate divisors (as classified in \cite{Carta:2022web}), can leave the axio-dilaton valley still flat. They are the most obvious pCICYs that may admit EFFV configurations. However, we noticed that this is not a sufficient condition, since for some of them, the symmetry is not respected at the level of intersection polynomial or the Gopakumar-Vafa (GV) invariants \cite{Gopakumar:1998ii,Gopakumar:1998jq} entering the (instanton) prepotential. After invoking a genuine symmetry considering all such exchange possibilities, we find only 13 pCICYs in total that can lead to a non-perturbative superpotential which gets identically trivial for such EFFV configurations. This investigation subsequently confirms that there are not too many CY 3-folds in which can result in EFFVs and hence physical vacua fixing the dilaton valley should be feasible in most of the pCICY 3-folds.

\end{itemize}

\noindent
The paper is organized as follows. In Section \ref{sec:analPFFV} we summarize the necessary steps to find perturbatively flat flux vacua (PFFV) as in \cite{Demirtas:2019sip,Carta:2021kpk}. We also start the systematic analysis of such PFFV in the pCICY manifolds with $h^{1,1}=2$ in Section \ref{sec:statPFFVCICYs}. In particular, in Section \ref{sec_K3-fibered} we discuss the PFFV for those mirror pCICYs that are K3-fibered, finding all possible physical vacua, and a similar analysis is done in Section \ref{sec_non-K3-fibered} for the mirror pCICYs that are not K3-fibered. In Section \ref{sec_EFFV} we focus on the EFFV. We conclude in Section \ref{sec:concl}. In Appendix \ref{App:pCICYgeom} we review the geometry of the pCICYs, and we collect all the data necessary to reproduce the computations.

\section{Analyzing the perturbatively flat flux vacua}
\label{sec:analPFFV}

In this section, first we briefly summarize the steps leading to perturbatively flat flux vacua introduced in \cite{Demirtas:2019sip}, which have been rephrased/recollected in \cite{Carta:2021kpk} subject to removing the flux scaling redundancies. Subsequently, we will use these steps to systematically characterize the perturbatively flat flux vacua for the mirror pCICY $3$-folds, which have recently received some significant amount of  phenomenological interests \cite{Carta:2021sms,Carta:2022web}. 

\subsection{Recipe for low $|W_0|$}
The overall recipe can be clubbed into the following four steps:
\begin{enumerate}[label={\bf S\arabic*}.,ref=S\arabic*]
\item\label{step1} 
{\bf Finding the flat vacua}: First, we look for a flux vector $M^i$ constructed using the three-form fluxes $\{F^i, H_i\}$ for which the following conditions hold:
\begin{enumerate}[label={\bf S1\alph*}.,ref=S1\alph*]
 \item\label{step1-N} There exists a matrix $N_{ij} = \kappa^0_{ijk}F^k$ such that $M^i=N^{ij}H_j$, which is orthogonal to the NS-NS flux vector, $H_i$, i.e. $M^iN_{ij}M^j=0$. Here the classical triple intersection numbers $\kappa_{ijk}^0$ are defined through integration using the two-form basis $\{J_i\}$ as below,
 \begin{equation}
 \kappa_{ijk}^0   =  \displaystyle{\int_{\tilde{X_3}} \, J_i \wedge J_j \wedge J_k}\coma
 \end{equation}
 
 \item\label{step1-Kahl} The flux vector $M^i$ must lie within the K\"ahler cone of the mirror CY. This means that in the $2$-form basis of the K\"ahler cone generators, then we demand $M^i>0$ for all the simplicial cases. 
\item\label{step1-tad} The flux vectors $F^i$ and $H_i$ must satisfy the tadpole constrains, i.e. $0\leq -F^iH_i\leq 2Q$, for a total D3-brane charge $Q$.
\item\label{step1-int} All the so-called shifted fluxes \cite{Blumenhagen:2014nba, Shukla:2019wfo} $F_i=\tilde{p}_{ij}F^j$ and $F_0=\tilde{p}_iF^i$ must be quantized to take integral values, where for a given CY $3$-fold $\tilde{X}_3$ \cite{Hosono:1994av,Arends:2014qca}, 
\begin{equation}
        \tilde{p}_{ij}  = \displaystyle{\frac{1}{2}\int_{\tilde{X_3}} \, J_i \wedge J_j \wedge J_j \quad ({\rm mod} \, \, {\mathbb Z})\coma} \tilde{p}_j  =  \displaystyle{\frac{1}{4\cdot 3!}\int_{\tilde{X_3}} \,c_2(\tilde{X_3}) \wedge J_j} \fstop 
        \end{equation}
    \end{enumerate}
\item\label{step2} 
{\bf Flux scaling redundancies}: In order to avoid several equivalent flux vacua related to each other by a simple scaling of the $F^i$ and $H_i$ fluxes, we demand \cite{Carta:2021kpk}
        \begin{equation}            
        \GCD\left(\{H_i\}\right) = 1\coma \GCD\left(\{F^i, \, H_i\}\right) = 1\fstop 
        \label{eq:gcd1}
        \end{equation}
\item\label{step3} 
{\bf Isolating PFFV with $M^1 = M^2$}: Applying Step \ref{step2}, we encounter PFFV configurations with $M^1=M^2$. It has been found in \cite{Carta:2021kpk} that such flux choices do not lead to trustworthy physical vacua after instanton corrections are taken into account through the prepotential. For these reasons, we demand the PFFV to satisfy Step \ref{step2} as well as $M^1 \neq M^2$. \\
However, depending on the presence of a symmetry in the CY geometry it might be possible that the non-perturbative effects are trivially decoupled in the EOM, leading to what we call as ``exponentially flat flux vacua". We will isolate such PFFV configuration for performing a separate analysis.
\item\label{step4}
{\bf Fixing the flat valley}: The K\"ahler potential of the complex structure moduli $(U^i)$ sector solely depends on the axio-dilaton $(S)$ along the perturbatively flat valley $U^i = S\, M^i$, and the ``effective" tree level axio-dilaton dependent term is given as \cite{Carta:2021kpk},
\begin{equation}
\label{eq:Keff}
K^{\rm eff}(S) \simeq - 4 \, \ln \left[-i \left(S -\ov{S} \right)\right] -\ln \left[\frac{1}{6} \kappa_{ijk}^0 \,M^i M^j M^k \right]\coma 
\end{equation}
where we have neglected the perturbative pieces involving the Euler characteristic of the mirror CY because that piece effectively appears with an extra factor of $g_s^{-3}$ and in the weak coupling regime, it is not likely to compete with the leading order effects. We define the axio-dilaton as $S = c_0 + i e^{-\phi}$ where $c_0$ is the universal axion and the string coupling $g_s$ is subsequently defined through the VEV of the dilaton as $g_s = \langle e^\phi \rangle$. Now, for stabilizing the perturbatively flat valley $U^i = S \, M^i$, we utilize the non-perturbative effects in the superpotential which takes the following form \cite{Carta:2021kpk},
\begin{equation}
\label{eq:Weff}
\begin{split}
W^{\rm eff}(S) \simeq &\, \frac{1}{(2 \pi)^2} \sqrt{\frac{2}{\pi}} \left[F^1 \, \left(n_1 \, e^{2\pi i S \, M^1} +2 \, n_{11} \,e^{4\pi i S \, M^1} + n_{12}\, e^{2\pi i S \, M^1}\, e^{2\pi i S\, M^2} \right)\right.+\\
& + F^2 \left.\left(n_2 \, e^{2\pi i S \, M^2} +2 \, n_{22} \,e^{4\pi i S \, M^2} + n_{12}\, e^{2\pi i S \, M^1}\, e^{2\pi i S\, M^2} \right) \right] + \ldots
\end{split}
\end{equation}
Here, $n_i$'s and $n_{ij}$ are defined as the coefficients appearing in the instanton  prepotential ${\cal F}_{\text{inst}}$ expressed in the following form,
\begin{equation}
\label{eq:F-inst-nis}
{\cal F}_{\text{inst}} = -\frac{1}{(2 \pi i)^3} \left(n_1 \, q_1 + n_2 \, q_2 + n_{11} \, q_1^2 + n_{12} \, q_1 \, q_2 + n_{22} \, q_2^2 +\ldots \right)\coma
\end{equation}
where $q_i = e^{2 \pi i\, U^i}$ for $i \in \{1, 2\}$ and we have considered only up to quadratic terms in $q_i$'s in the instanton expansion. These quantities $n_i$ and $n_{ij}$ are collected for all the 36 pCICY geometries in Table \ref{tab:nijh112}. Subsequently, the flat vacua are stabilized by giving the dilaton a VEV by solving the following supersymmetric flatness condition:
\begin{equation}
\label{eq:SUSY-eff}
D_S W^{\rm eff}(S) = \partial_S W^{\rm eff}(S) + W^{\rm eff}(S) (\partial_S K^{\rm eff}) = 0\fstop
\end{equation}
In addition, we demand that for the vacua to be physical in the sense of lying within the weak string-coupling and large complex-structure regime, i.e., $\langle g_s \rangle < 1, \langle u^i \rangle > 1$ and $\langle u^2 \rangle > 1$.
\end{enumerate}

\subsection{Demonstrating the steps in an explicit example}
\label{sec:explexampl}

In this section, we present an explicit example to illustrate the various steps. Subsequently, the same strategy will be replicated with all the remaining 35 CYs. For this purpose, we take the pCICY called as $\mathcal{M}_{2,33}$ in Table \ref{tab_data-h11=2} where the relevant topological data for this pCICY can also be found. The task of scanning PFFV demands to find flux configurations satisfying the following explicit conditions,
\begin{equation}
    \label{eq:final-constraints}
    \begin{dcases}
   M^1 = -\frac{H_2}{4\, F^2} > 0\coma\\
   M^2 = \frac{H_1}{4\, F^2} > 0\coma \\
    N_{\rm flux} = \frac18 F^2 \left(5 H_1 - 12 H_2 \right) \in {\mathbb Z} & \text{such that }0 < N_{\rm flux} \leq N_{\rm flux}^{\text{max}}\coma\\
   \tilde{p}_{ij}F^j = \left\{0, \, \, \frac52 F^2 \right\} \in {\mathbb Z} \times {\mathbb Z}\coma\\
   \tilde{p}_i F^i = F^2 \left(\frac56 + \frac{2 H_2}{H_1} \right) \in {\mathbb Z} \coma \\
    F^1 = F^2 \left(\frac{2H_2}{H_1} -\frac54 \right) \in {\mathbb Z}\fstop
    \end{dcases}
\end{equation}
Here the last condition follows from solving the orthogonality relation $M^i \, H_i = 0$ for the $F^1$ flux, which has been subsequently eliminated from the other ingredients such as $M^i$, $\tilde{p}_{ij}F^j$,  $\tilde{p}_i F^i$ and $N_{\rm flux}$ keeping them as a function of only three fluxes, namely $F^2, H_1$ and $H_2$. As we observed in \cite{Carta:2021kpk}, the conditions listed in Step \ref{step1} for the pCICYs with $h^{1,1} = 2$ define two possible choice of fluxes, i.e.
\begin{equation}
\label{eq:flux-choice}
\{H_1 > 0,\,H_2 < 0,\, F^2 > 0\} \quad \text{or} \quad \{H_1 < 0,\, H_2 > 0,\, F^2 < 0\}\fstop
\end{equation}
This means that also for the pCICY database, we can continue to scan among all the possible flux configurations defined in the following region in the flux space 
\begin{equation}
H_1 \in \{-300, -1\}\coma \quad H_2 \in \{1, 300\}\coma \quad F^2 \in \{-300, -1\}\coma
\end{equation}
which has to satisfy Step \ref{step1}. Given that the K\"aher cone conditions are already imposed on the flux vector $M^i$, the intermediate quantities are not expected to get singular within this choice of range for the fluxes.

We find that there are a total of 41 flux configurations which satisfy the conditions of being PFFV as mentioned in Step \ref{step1}. Subsequently, following the Step \ref{step2} about the flux scaling redundancy, this number reduces into 26 while further imposing the condition $M^1 \neq M^2$ in Step \ref{step3}, one is left with 24 flux configurations as presented in Table \ref{tab:41vacuaNflux150}. Subsequently, after performing Step \ref{step4}, we note that there are only 2 flux configurations (numbered as 8 and 20 in the collection of 24 PFFV in the Table \ref{tab:41vacuaNflux150}) which result in physical VEVs for dilaton and complex structure moduli. The details on these physical various along with the moduli/dilaton VEVs are given respectively as below,

\begin{equation}
 \begin{array}{rllrllrll}
   F^i                      & = & \{33, -12\}\coma & H_i                 & = & \{-4, 3\}\coma              & M^i          & = & \{1/16, \, 1/12\}\coma\\
   \tilde{p}_{ij} F^j       & = & \{0, -30\}\coma  & \, \, \tilde{p}_i F^i     & = & 8\coma                     & N_{\rm flux} & = & 84\coma\\
   \langle g_s \rangle^{-1} & = & 22.2784 \coma     & \langle U^i \rangle & = & \{1.3924, 1.85653\} \coma  & |W_0|        & = & 4.94604 \cdot 10^{-6}\coma\\
\end{array}  
\end{equation}
and
\begin{equation}
 \begin{array}{rllrllrll}
   F^i                      & = & \{35, -12\}\coma & H_i                 & = & \{-6, 5\}\coma              & M^i          & = & \{5/48, \, 1/8\}\coma\\
   \tilde{p}_{ij} F^j       & = & \{0, -30\}\coma  & \, \, \tilde{p}_i F^i     & = & 10\coma                     & N_{\rm flux} & = & 135\coma\\
   \langle g_s \rangle^{-1} & = & 21.2343\coma     & \langle U^i \rangle & = & \{2.21191, 2.65429\} \coma  & |W_0|        & = & 1.231 \cdot 10^{-8}\fstop\\
\end{array}  
\end{equation}
\noindent
In fact the PFFV numbered as 15  (out of the list of 24) in the Table \ref{tab:41vacuaNflux150} gives a quite small value of the string coupling $g_s$, however, one of the complex-structure saxion slightly goes beyond the physical regime as seen below:
\begin{equation}
 \begin{array}{rllrllrll}
   F^i                      & = & \{62, -24\}\coma & H_i                 & = & \{-3, 2\}\coma              & M^i          & = & \{1/48, \, 1/32\}\coma\\
   \tilde{p}_{ij} F^j       & = & \{0, -60\}\coma  & \, \, \tilde{p}_i F^i     & = & 12\coma                     & N_{\rm flux} & = & 117\coma\\
   \langle g_s \rangle^{-1} & = & 46.5371 \coma     & \langle U^i \rangle & = & \{0.969522, 1.45428\} \coma  & |W_0|        & = & 1.74145 \cdot 10^{-4}\fstop\\
\end{array}  
\end{equation}

\begin{center}
\renewcommand*{\arraystretch}{1.2}
\begin{longtable}{|c||c|cc|cc|cc|cc|c|}
\caption{Details of 24 PFFV obtained after Step \ref{step3}, and by demanding $0 < N_{\rm flux} \leq 150$.}\\
\hline
 PFFV \# &  $N_{\rm flux} $& $H_1 $& $H_2  $& $F^1 $& $F^2$&  $M^1  $& $M^2  $& $\tilde{p}_{1j}F^j $& $\tilde{p}_{2j}F^j $& $\tilde{p}_{i}F^i$ \\
 \hhline{|=#=|==|==|==|==|=|}
\endfirsthead
 \hline
 PFFV \# &  $N_{\rm flux} $& $H_1 $& $H_2  $& $F^1 $& $F^2$&  $M^1  $& $M^2  $& $\tilde{p}_{1j}F^j $& $\tilde{p}_{2j}F^j $& $\tilde{p}_{i}F^i$ \\
 \hhline{|=#=|==|==|==|==|=|}
 \endhead
 \label{tab:41vacuaNflux150}
 $1  $&  $33  $& $-2  $& $1 $& $27  $& $-12 $& $\frac{1}{48}$  & $\frac{1}{24}$  & $0$ & $-30  $& $2   $ \\
 $2  $&  $48  $& $-4  $& $1 $& $21  $& $-12 $& $\frac{1}{48}$  & $\frac{1}{12}$  & $0$ & $-30  $& $-4  $ \\
 $3  $&  $63  $& $-6  $& $1 $& $19  $& $-12 $& $\frac{1}{48}$  & $\frac{1}{8}$   & $0$ & $-30  $& $-6  $ \\
 $4  $&  $66  $& $-2  $& $1 $& $54  $& $-24 $& $\frac{1}{96}$  & $\frac{1}{48}$  & $0$ & $-60  $& $4   $ \\
 $5  $&  $69  $& $-2  $& $3 $& $51  $& $-12 $& $\frac{1}{16}$  & $\frac{1}{24}$  & $0$ & $-30  $& $26  $ \\
 $6  $&  $78  $& $-8  $& $1 $& $18  $& $-12 $& $\frac{1}{48}$  & $\frac{1}{6}$   & $0$ & $-30  $& $-7  $ \\
 $7 $&  $81  $& $-3  $& $1 $& $46  $& $-24 $& $\frac{1}{96}$  & $\frac{1}{32}$  & $0$ & $-60  $& $-4  $ \\
 $8 $&  $84  $& $-4  $& $3 $& $33  $& $-12 $& $\frac{1}{16}$  & $\frac{1}{12}$  & $0$ & $-30  $& $8   $ \\
 $9 $&  $87  $& $-1  $& $2 $& $126 $& $-24 $& $\frac{1}{48}$  & $\frac{1}{96}$  & $0$ & $-60  $& $76  $ \\
 $10 $&  $96  $& $-4  $& $1 $& $42  $& $-24 $& $\frac{1}{96}$  & $\frac{1}{24}$  & $0$ & $-60  $& $-8  $ \\
 $11 $&  $99  $& $-2  $& $1 $& $81  $& $-36 $& $\frac{1}{144}$ & $\frac{1}{72}$  & $0$ & $-90  $& $6   $ \\
 $12 $&  $105 $& $-2  $& $5 $& $75  $& $-12 $& $\frac{5}{48}$  & $\frac{1}{24}$  & $0$ & $-30  $& $50  $ \\
 $13 $&  $108 $& $-12 $& $1 $& $17  $& $-12 $& $\frac{1}{48}$  & $\frac{1}{4}$   & $0$ & $-30  $& $-8  $ \\
 $14 $&  $114 $& $-8  $& $3 $& $24  $& $-12 $& $\frac{1}{16}$  & $\frac{1}{6}$   & $0$ & $-30  $& $-1  $ \\
 $15 $&  $117 $& $-3  $& $2 $& $62  $& $-24 $& $\frac{1}{48}$  & $\frac{1}{32}$  & $0$ & $-60  $& $12  $ \\
 $16 $&  $120 $& $-4  $& $5 $& $45  $& $-12 $& $\frac{5}{48}$  & $\frac{1}{12}$  & $0$ & $-30  $& $20  $ \\
 $17 $&  $123 $& $-1  $& $3 $& $174 $& $-24 $& $\frac{1}{32}$  & $\frac{1}{96}$  & $0$ & $-60  $& $124 $ \\
 $18 $&  $126 $& $-6  $& $1 $& $38  $& $-24 $& $\frac{1}{96}$  & $\frac{1}{16}$  & $0$ & $-60  $& $-12 $ \\
 $19 $&  $132 $& $-2  $& $1 $& $108 $& $-48 $& $\frac{1}{192}$ & $\frac{1}{96}$  & $0$ & $-120 $& $8   $ \\
 $20 $&  $135 $& $-6  $& $5 $& $35  $& $-12 $& $\frac{5}{48}$  & $\frac{1}{8}$   & $0$ & $-30  $& $10  $ \\
 $21 $&  $138 $& $-2  $& $3 $& $102 $& $-24 $& $\frac{1}{32}$  & $\frac{1}{48}$  & $0$ & $-60  $& $52  $ \\
 $22 $&  $141 $& $-2  $& $7 $& $99  $& $-12 $& $\frac{7}{48}$  & $\frac{1}{24}$  & $0$ & $-30  $& $74  $ \\
 $23 $&  $144 $& $-4  $& $1 $& $63  $& $-36 $& $\frac{1}{144}$ & $\frac{1}{36}$  & $0$ & $-90  $& $-12 $ \\
 $24 $&  $150 $& $-8  $& $5 $& $30  $& $-12 $& $\frac{5}{48}$  & $\frac{1}{6}$   & $0$ & $-30  $& $5   $ \\
\hline
\end{longtable}     
\end{center}

\section{Statistics of perturbative flux vacua for pCICYs}
\label{sec:statPFFVCICYs} 

Following the methodology described for the explicit example in the previous section, we performed the flux vacua analysis for all the 36 pCICYs with $h^{1,1} =2$, and the results obtained by following the three steps \ref{step1}-\ref{step3} are shown in Table \ref{tab_fluxdata}. We also note that the number of flat flux vacua listed in Table \ref{tab_fluxdata} are obtained by choosing the three independent fluxes $|F^2|, |H_1|, |H_2|$ in the range $\{1, 300\}$ while the remaining flux $F^1$ is obtained by solving the orthogonality condition $M^i H_i  = 0$. 

\begin{center}
\renewcommand\arraystretch{1}
   \begin{longtable}{|c|c|c||c|c|c|}
\caption{Number of PFFV obtained after each of the three steps \ref{step1}-\ref{step3}, corresponding to $N_{\rm flux} \leq 500$. The second column is the number of the pCICY as in \cite{Anderson:2017aux}. From now on, we will omit this number, and we refer to the pCICY using the first column.} \\
\hline
CY & pCICY & K3-fibred &\multicolumn{3}{c|}{\#(PFFV) for $N_{\rm flux} \leq 500$} \\
\cline{4-6}
 \# & \# & & Step \ref{step1} & Step \ref{step2} & Step \ref{step3}  \\
\hhline{|=|=|=|=#=|=|}
\endfirsthead
\hline
CY & pCICY & K3-fibred &\multicolumn{3}{c|}{\#(PFFV) for $N_{\rm flux} \leq 500$} \\
\cline{4-6}
 \# & \# & & Step \ref{step1} & Step \ref{step2} & Step \ref{step3}  \\
\hhline{|=|=|=|=#=|=|}
\endhead
\label{tab_fluxdata}
 1 & 7643 & no & 254 & 100 & 90 \\
 2 & 7644 & no & 2913 & 349 & 49 \\
 3 & 7668 & no & 467 & 156 & 126 \\
 4 & 7725 & no & 1443 & 309 & 159 \\
 5 & 7726 & no & 221 & 57 & 41 \\
 6 & 7758 & no & 114 & 47 & 31 \\
 7 & 7759 & no & 505 & 118 & 97 \\
 8 & 7761 & no & 1348 & 167 & 17 \\
 9 & 7799 & no & 1348 & 165 & 15 \\
 10 & 7806 & yes & 10432 & 3019 & 2769 \\
 11 & 7807 & no & 214 & 62 & 46 \\
 12 & 7808 & no & 580 & 164 & 89 \\
 13 & 7809 & no & 27 & 13 & 8 \\
 14 & 7816 & yes & 3058 & 1026 & 943 \\
 15 & 7817 & yes & 4150 & 1397 & 1342 \\
 16 & 7819 & yes & 2276 & 811 & 778 \\
 17 & 7821 & no & 407 & 124 & 121 \\
 18 & 7822 & yes & 3058 & 1026 & 943 \\
 19 & 7823 & yes & 2276 & 811 & 778 \\
 20 & 7833 & no & 108 & 47 & 44 \\
 21 & 7840 & yes & 2066 & 889 & 834 \\
 22 & 7844 & no & 340 & 102 & 88 \\
 23 & 7853 & no & 312 & 105 & 80 \\
 24 & 7858 & yes & 655 & 325 & 304 \\
 25 & 7863 & no & 2913 & 349 & 49 \\
 26 & 7867 & yes & 4810 & 1872 & 1772 \\
 27 & 7868 & no & 28 & 17 & 16 \\
 28 & 7869 & yes & 4810 & 1872 & 1772 \\
 29 & 7873 & yes & 2683 & 920 & 882 \\
 30 & 7882 & yes & 1925 & 817 & 772 \\
 31 & 7883 & no & 263 & 87 & 83 \\
 32 & 7884 & no & 1905 & 348 & 198 \\
 33 & 7885 & yes & 411 & 192 & 183 \\
 34 & 7886 & yes & 2276 & 811 & 778 \\
 35 & 7887 & yes & 1648 & 610 & 587 \\
 36 & 7888 & yes & 2276 & 811 & 778 \\
 \hline 
\end{longtable}
\end{center}

\noindent
Note that the CY \# 32 in Table \ref{tab_fluxdata} is the same which appears in the Kreuzer-Skarke database as well, and therefore its statistics matches with the one presented in \cite{Carta:2021kpk}.

From Table \ref{tab_fluxdata} we also observe that although Step \ref{step1} results in a reasonably large number of PFFV for most of the examples, applying the next steps \ref{step2} and \ref{step3} reduces these numbers vary significantly. For example, the K3-fibered CY numbered as 10 of Table \ref{tab_fluxdata} results a total of 10432 number of PFFV from Step \ref{step1} which is reduced to 3019 after Step \ref{step2}, and is further reduced to 2769 after Step \ref{step3}.

\subsubsection*{Some comments on PFFV with $M^1 =M^2$}
In order to appreciate the reduction in the number of PFFV in Step \ref{step3}, let us present some insights on this step. Actually, the PFFV configurations with $M^1 = M^2$ are equivalent to $H_1 + H_2 = 0$ and do not result in trustworthy physical vacua through Step \ref{step4}. As observed in \cite{Carta:2021kpk}, this happens because the dilaton minimization process uses two pieces appearing at different orders of suppression within the same series expansion of the non-perturbative effects. To see it quickly, one can consider the following simplified form of the supersymmetric flatness condition \cref{eq:SUSY-eff} which is obtained by using $M^1 = M^2 = M$,
\begin{equation}
\label{eq:SUSY-eff-M1=M2}
\begin{split}
e^{2 \pi  {M} \,\langle s \rangle } \, {(\pi  {M} s+1)\, ({F^1} {n_1}+{F^2} {n_2})}  = -(2 \pi  {M} s+1)\, ({F^1} (2 {n_{11}}+{n_{12}})+{F^2} ({n_{12}}+2 {n_{22}})) \coma
\end{split}
\end{equation}
where $s={\rm Im}(S)$. This shows that the first order  non-perturbative terms will have to complete with the second order terms, and neglecting the pieces containing $n_{ij}$'s lead to no physical solution. Therefore, for all the PFFV configurations resulting in $M^1 = M^2$ one can check the sign of the following quantity,
\begin{equation}
\label{eq:calA}
{\cal A} = \frac{{F^1} (2 {n_{11}}+{n_{12}})+{F^2} ({n_{12}}+2 {n_{22}})}{{F^1} {n_1}+{F^2} {n_2}}\fstop
\end{equation}
 If one finds that ${\cal A} \geq 0$, this would imply that there is no physical solution in the weak coupling regime for the corresponding flat vacua. Also, even if ${\cal A} < 0$, one has to check that one gets weak coupling i.e., $ \langle s \rangle > 1$. By these criteria, one can discard several flat vacua arising from Step \ref{step1} and Step \ref{step2} which is analyzed in Table \ref{tab:M1M2values} as a result of Step \ref{step3}. 

\begin{center}
\renewcommand{\arraystretch}{1.15}
  \begin{longtable}{|c|cccc||c|cccc|}
  \caption{Number of PFFV configurations with $M^1 =M^2 = a \, H_1/{F^2}$ along with the sign($\cal A$) where ${\cal A}$ is defined in Eq. \eqref{eq:calA}. There are $5$ cases for which the parameter $\mathcal{A}$ turns out to be of the form $\frac00$ which we will explore separately.
  }\\
    \hline
CY \# &$a$ & $\#(M^1 = M^2)$ & ${\cal A} > 0$ & ${\cal A} < 0$ & CY \# &$a$ & $M^1 = M^2$ & ${\cal A} > 0$ & ${\cal A} < 0$ \\   
\hhline{|=|====#=|====|}
\endfirsthead
  \hline
CY \# &$a$ & $\#(M^1 = M^2)$ & ${\cal A} > 0$ & ${\cal A} < 0$ & CY \# &$a$ & $M^1 = M^2$ & ${\cal A} > 0$ & ${\cal A} < 0$ \\   
\hhline{|=|====#=|====|}
  \endhead
  \label{tab:M1M2values}$1 $ & $\frac{5}{44} $& $28   $& $0    $& $28  $& $19$ & $\frac{1}{8}   $& $196  $& $196 $& $0  $ \\
$2 $ & $\frac{1}{8}  $& $2790 $& $-    $& $-   $& $20$ & $\frac{13}{73} $& $8    $& $0   $& $8  $  \\
 $3 $ & $\frac{5}{33} $& $126  $& $0    $& $126 $& $21$ & $\frac{1}{6}   $& $231  $& $231 $& $0  $ \\
 $4 $ & $\frac{1}{8}  $& $860  $& $0    $& $860 $& $22$ & $\frac{7}{34}  $& $60   $& $0   $& $60 $  \\
 $5 $ & $\frac{3}{20} $& $116  $& $0    $& $116 $& $23$ & $\frac{1}{6}   $& $115  $& $0   $& $115$ \\
 $6 $ & $\frac{9}{70} $& $65   $& $0    $& $65  $& $24$ & $\frac{1}{6}   $& $70   $& $70  $& $0  $  \\
 $7 $ & $\frac{7}{46} $& $111  $& $111  $& $0   $& $25$ & $\frac{1}{4}   $& $2790 $& $-   $& $-  $ \\
 $8 $ & $\frac{1}{5}  $& $1321 $& $-    $& $-   $& $26$ & $\frac{1}{6}   $& $482  $& $482 $& $0  $  \\
 $9 $ & $\frac{1}{5}  $& $1321 $& $-    $& $-   $& $27$ & $\frac{13}{64} $& $2    $& $0   $& $2  $ \\
 $10$ & $\frac{1}{6}  $& $1421 $& $1421 $& $0   $& $28$ & $\frac{1}{6}   $& $482  $& $482 $& $0  $  \\
 $11$ & $\frac{9}{62} $& $101  $& $0    $& $101 $& $29$ & $\frac{1}{6}   $& $232  $& $232 $& $0  $ \\
 $12$ & $\frac{1}{6}  $& $371  $& $0    $& $371 $& $30$ & $\frac{1}{6}   $& $182  $& $182 $& $0  $  \\
 $13$ & $\frac{5}{26} $& $14   $& $0    $& $14  $& $31$ & $\frac{11}{43} $& $12   $& $0   $& $12 $ \\
 $14$ & $\frac{1}{8}  $& $379  $& $379  $& $0   $& $32$ & $\frac{1}{3}   $& $1321 $& $-   $& $-  $  \\
 $15$ & $\frac{1}{8}  $& $367  $& $367  $& $0   $& $33$ & $\frac{1}{4}   $& $38   $& $38  $& $0  $ \\
 $16$ & $\frac{1}{8}  $& $196  $& $196  $& $0   $& $34$ & $\frac{1}{4}   $& $196  $& $196 $& $0  $  \\
 $17$ & $\frac{4}{23} $& $7    $& $0    $& $7   $& $35$ & $\frac{1}{4}   $& $123  $& $123 $& $0  $ \\
 $18$ & $\frac{1}{8}  $& $379  $& $379  $& $0   $& $36$ & $\frac{1}{4}   $& $196  $& $196 $& $0  $  \\
 \hline
  \end{longtable}
\end{center}

\noindent
In Table \ref{tab:M1M2values}, we present some estimates about the $\sign({\cal A})$ for all such vacua in each of the pCICY $3$-fold, which shows that it is always fixed in each of the CY $3$-fold. Moreover, ${\cal A}$ takes an undetermined form for five pCICY examples and let us note that in the presence of a symmetry in the CY $3$-fold, such PFFV with $M^1 = M^2$ get interesting in the sense that they remain flat against the non-perturbative effects as well, and we will discuss such a class of vacua in the upcoming section.

\subsection{K3-fibered mirror pCICYs}
\label{sec_K3-fibered}

\begin{figure}[!htp]
 \centering
      \pgfplotstableread{
Label N50 N100 N150 N200 N250 N300 N350 N400 N450 N500 topper
10  348 679 843 1005 1070 1202 1244 1335 1379 1327 0 
14  71 171 243 280 297 382 336 411 457 410 0
15  104 241 333 381 389 529 456 581 609 527 0
16  42 117 173 203 222 284 253 315 347 320 0
18  71 171 243 280 297 382 336 411 457 410 0
19  42 117 173 203 222 284 253 315 347 320 0
21  43 111 163 183 194 271 229 299 310 263 0
24  8 30 42 60 63 80 83 90 91 108 0
26  126 282 363 449 488 547 579 627 654 695 0
28  126 282 363 449 488 547 579 627 654 695 0
29  54 146 201 240 277 312 325 350 374 404 0
30  39 104 144 173 196 226 233 250 269 291 0
33  2 16 23 38 40 46 55 58 63 70 0
34  42 117 173 203 222 284 253 315 347 320 0
35  32 86 128 147 159 213 181 230 250 222 0
36  42 117 173 203 222 284 253 315 347 320 0
    }\testdata
         \begin{tikzpicture}[scale=1]
    \begin{axis}[
       width=0.8\textwidth,
        ybar stacked,
        ymin=0,
        ymax=11000,
        xtick={data},
        legend style={cells={anchor=west}, legend pos=north east},
        reverse legend=false, 
        xticklabels from table={\testdata}{Label},
        xticklabel style={text width=2cm,align=center},
      bar width=0.032\textwidth,
      xlabel={CY \#},
		ylabel={\# flat vacua},
    ]
    \addplot [fill=col1,ybar]
            table [y=N50, meta=Label, x expr=\coordindex]
                {\testdata};
                    \addlegendentry{50}
        \addplot [fill=col2,ybar]
            table [y=N100, meta=Label, x expr=\coordindex]
                {\testdata};
                    \addlegendentry{100}
        \addplot [fill=col3]
            table [y=N150, meta=Label, x expr=\coordindex]
                {\testdata};
                    \addlegendentry{150}
    \addplot [fill=col4,ybar]
            table [y=N200, meta=Label, x expr=\coordindex]
                {\testdata};
                    \addlegendentry{200}
        \addplot [fill=col5,ybar]
            table [y=N250, meta=Label, x expr=\coordindex]
                {\testdata};
                    \addlegendentry{250}
        \addplot [fill=col6]
            table [y=N300, meta=Label, x expr=\coordindex]
                {\testdata};
                    \addlegendentry{300}
    \addplot [fill=col7,ybar]
            table [y=N350, meta=Label, x expr=\coordindex]
                {\testdata};
                    \addlegendentry{350}
        \addplot [fill=col8,ybar]
            table [y=N400, meta=Label, x expr=\coordindex]
                {\testdata};
                    \addlegendentry{400}
        \addplot [fill=col9]
            table [y=N450, meta=Label, x expr=\coordindex]
                {\testdata};
                    \addlegendentry{450}
    \addplot [fill=col10
    ]
            table [y=N500, meta=Label, x expr=\coordindex]
                {\testdata};
                    \addlegendentry{500}
    \end{axis}
    \end{tikzpicture}
\caption{Statistics of PFFV for K3-fibered models. Here different colors presenting the PFFV numbers for each of the CY space correspond to $N_{\rm flux} \leq 50 \, n$ where $n = \{1, 2, .., 10\}$.}
\label{fig:PFFV-fibered}
\end{figure}
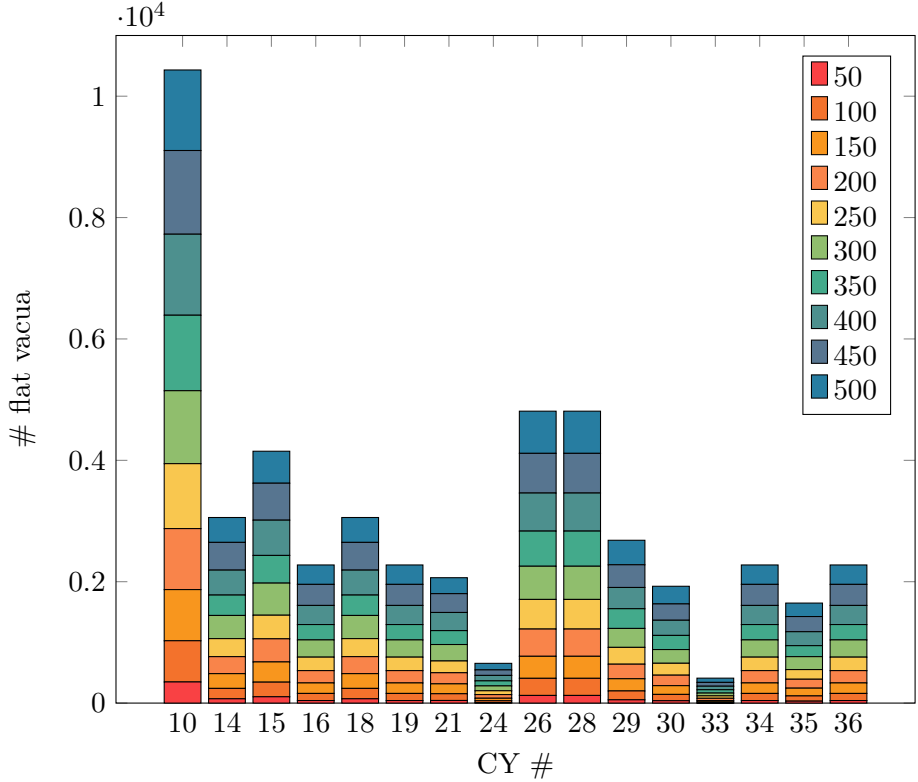

For pCICY $3$-fold with $h^{1,1}=2$, there are 16 geometries which are K3-fibered. These are in Table \ref{tab_fluxdata} where we present the corresponding number of PFFV configurations for $N_{\rm flux} \leq 500$. Let us mention that there are a total of 48810 PFFV configurations arising through the K3-fibered pCICYs which is a significant percentage of the total number of PFFV being 64520 obtained after Step \ref{step1}. The statistical distribution of these 48810 PFFV is shown in Figure \ref{fig:PFFV-fibered} with a distinction based on a range of $D3$ tadpole charge given as $0 < N_{\rm flux} \leq 50 \, n$ where $n = \{1, 2, .., 10\}$.

\subsubsection*{Finding the physical vacua}
In fact, in order to give some more insights at the intermediate steps, let us mention that although there is a total of 48810 number of flat vacua arising from the 16 K3-fibered pCICYs after Step \ref{step1}, using our prescription described in three steps \ref{step1}-\ref{step3}, this number is reduced to 17209 after Step \ref{step2}, and finally to 16215 after the Step \ref{step3}. We investigate all these PFFV configurations for fixing the dilaton by using the non-perturbative effects in Step \ref{step4}, and find that there remain only 559 PFFV which give physical solutions in the sense of being in the weak coupling and large complex structure limit. To be more specific, we impose: $\langle g_s \rangle < 1, \langle u^1 \rangle > 1$ and $\langle u^2 \rangle > 1$, however, this appears to be quite restrictive for many of the PFFV to be ruled out. Although we provide the details of all the 559 physical vacua in a separate attachment for the interested readers, for illustration purpose here we present one such physical vacua with the lowest value of $|W_0|\simeq 2.0769 \cdot 10^{-27}$ as below,
\begin{eqnarray}
\label{eq:}
& & F^i = \{39, -10\}\coma \quad H_i =  \{-20, 19\}\coma \quad M^i = \{19/60, \, 1/3\}\coma \\
& &    \tilde{p}_{ij} F^j  = \{0, -60\}\coma \quad \, \, \tilde{p}_i F^i = 14\coma \quad N_{\rm flux} = 485\coma \quad \langle K(s, u^i) \rangle = -14.1755 \coma \nonumber\\
& & \langle g_s \rangle^{-1} = 26.5741 \coma \quad \langle U^i \rangle = \{8.41514, \, 8.85805\} \coma \quad |W_0| = 2.0769 \cdot 10^{-27}\fstop \nonumber
\end{eqnarray}

\noindent
Finally let us mention that all the 16 K3-fibered pCICYs can indeed give physical vacua via fixing the axio-dilaton flat valley through the non-perturbative effects in ${\cal F}_{\rm inst}$.

\subsection{Non K3-fibered mirror pCICYs}
\label{sec_non-K3-fibered}

\begin{figure}[!htp]
 \centering
      \pgfplotstableread{
Label N50 N100 N150 N200 N250 N300 N350 N400 N450 N500 topper
 1 6 14 25 23 22 43 28 38 34 21 0  
 2 207 281 308 330 333 366 253 286 278 271 0  
 3 13 33 39 49 42 64 61 59 62 45 0  
 4 68 117 124 155 141 173 152 193 153 167 0  
 5 8 13 24 18 21 31 29 28 28 21 0  
 6 3 6 9 12 9 16 17 17 12 13 0  
 7 25 40 51 48 52 67 52 60 68 42 0  
 8 87 122 131 148 141 165 130 148 141 135 0  
 9 87 120 135 144 145 161 132 140 149 135 0  
11 7 13 17 19 21 26 27 25 36 23 0  
12 24 40 52 60 57 64 69 82 63 69 0  
13 0 1 3 3 2 3 6 5 0 4 0  
17 14 29 37 36 46 46 37 57 58 47 0  
20 1 7 8 11 11 17 2 20 19 12 0  
22 10 22 34 28 34 55 34 46 48 29 0  
23 10 20 28 27 34 39 33 34 52 35 0  
25 207 281 308 330 333 366 253 286 278 271 0  
27 0 1 1 2 3 5 3 5 3 5 0  
31 9 17 23 23 29 31 33 30 30 38 0  
32 101 160 185 202 197 245 185 234 216 180 0 
    }\testdata
         \begin{tikzpicture}[scale=1]
    \begin{axis}[
       width=\textwidth,
        ybar stacked,
        ymin=0,
        ymax=3800,
        xtick={data},
        legend style={cells={anchor=west}, legend pos=north east},
        reverse legend=false, 
        xticklabels from table={\testdata}{Label},
        xticklabel style={text width=2cm,align=center},
      bar width=0.032\textwidth,
      xlabel={CY \#},
		ylabel={\# flat vacua},
    ]
    \addplot [fill=col1,ybar]
            table [y=N50, meta=Label, x expr=\coordindex]
                {\testdata};
                    \addlegendentry{50}
        \addplot [fill=col2,ybar]
            table [y=N100, meta=Label, x expr=\coordindex]
                {\testdata};
                    \addlegendentry{100}
        \addplot [fill=col3]
            table [y=N150, meta=Label, x expr=\coordindex]
                {\testdata};
                    \addlegendentry{150}
    \addplot [fill=col4,ybar]
            table [y=N200, meta=Label, x expr=\coordindex]
                {\testdata};
                    \addlegendentry{200}
        \addplot [fill=col5,ybar]
            table [y=N250, meta=Label, x expr=\coordindex]
                {\testdata};
                    \addlegendentry{250}
        \addplot [fill=col6]
            table [y=N300, meta=Label, x expr=\coordindex]
                {\testdata};
                    \addlegendentry{300}
    \addplot [fill=col7,ybar]
            table [y=N350, meta=Label, x expr=\coordindex]
                {\testdata};
                    \addlegendentry{350}
        \addplot [fill=col8,ybar]
            table [y=N400, meta=Label, x expr=\coordindex]
                {\testdata};
                    \addlegendentry{400}
        \addplot [fill=col9]
            table [y=N450, meta=Label, x expr=\coordindex]
                {\testdata};
                    \addlegendentry{450}
    \addplot [fill=col10
    ]
            table [y=N500, meta=Label, x expr=\coordindex]
                {\testdata};
                    \addlegendentry{500}
    \end{axis}
    \end{tikzpicture}
\caption{Statistics of PFFV for non K3-fibered CYs. Here, different colors presenting the PFFV numbers for each of the CY space correspond to $N_{\rm flux} \leq 50 \, n$ where $n = \{1, 2, .., 10\}$.}
\label{fig:PFFVremainmodel}
\end{figure}
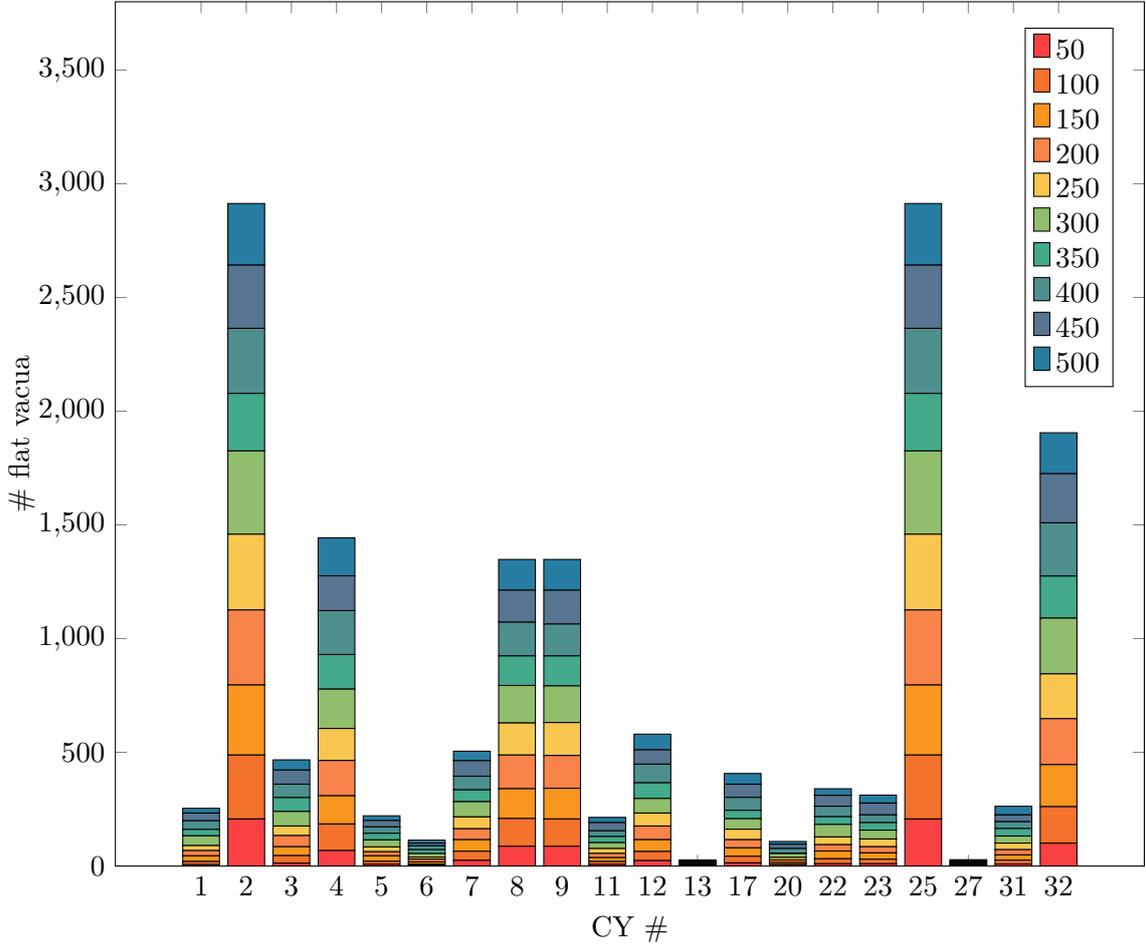

For pCICY $3$-folds with $h^{1,1}=2$, there are 20 geometries which are not of the K3-fibered type as mentioned in Table \ref{tab_fluxdata}, where we have also presented the number of PFFV configurations for $N_{\rm flux} \leq 500$. We find that there is a reasonably good number  of the PFFV in this class to begin with, i.e., after Step \ref{step1}, which results in 15710 PFFV configurations, with their statistical distributions being shown in Figure \ref{fig:PFFVremainmodel}. However, these numbers reduce significantly to 2886 after Step \ref{step2}, and to 1447 after Step \ref{step3}. This illustrates the fact that our three-step strategy \ref{step1}-\ref{step3} turns out to be crucially useful in eliminating the over-counting of (equivalent) vacua, before looking for the axio-dilaton stabilization using non-perturbative effects. 

\noindent
\subsubsection*{Finding the physical vacua}
After considering the axio-dilaton minimization along with demanding the weak coupling and large complex-structure VEV requirements for all the 1447 PFFV configurations remained after Step \ref{step3}, one is left with only 3 physical vacua arising from 2 out of 20 non K3-fibred pCICYs. These are numbered as 3 and 12 in the collection of 36 pCICYs in Table \ref{tab_fluxdata}. We find that the CY \# 3 has a unique physical solution described by the following data,
\begin{equation}
 \begin{array}{rllrllrll}
   F^i                      & = & \{54, -32\}\coma & H_i                 & = & \{-8, 7\}\coma              & M^i          & = & \{1/30, \, 4/105\}\coma\\
   \tilde{p}_{ij} F^j       & = & \{-48, -15\}\coma  & \, \, \tilde{p}_i F^i     & = & 17\coma                     & N_{\rm flux} & = & 328\coma\\
   \langle g_s \rangle^{-1} & = & 53.6791\coma     & \langle U^i \rangle & = & \{1.7893, 2.04492\} \coma  & |W_0|        & = & 2.19478 \cdot 10^{-7}\fstop\\
\end{array}  
\end{equation}
Further, the CY \# 12 has the following two physical vacua which have some correlations,
\begin{equation}
 \begin{array}{rllrllrll}
   F^i                      & = & \{86, -48\}\coma & H_i                 & = & \{-4, 3\}\coma              & M^i          & = & \{1/94, \, 2/141\}\coma\\
   \tilde{p}_{ij} F^j       & = & \{-72, -87\}\coma  & \, \, \tilde{p}_i F^i     & = & 21\coma                     & N_{\rm flux} & = & 244\coma\\
   \langle g_s \rangle^{-1} & = & 95.22\coma     & \langle U^i \rangle & = & \{1.01298, 1.35064\} \coma  & |W_0|        & = & 7.37045 \cdot 10^{-5}\coma\\
\end{array}  
\end{equation}
and
\begin{equation}
 \begin{array}{rllrllrll}
   F^i                      & = & \{172, -96\}\coma & H_i                 & = & \{-4, 3\}\coma              & M^i          & = & \{1/188, \, 1/141\}\coma\\
   \tilde{p}_{ij} F^j       & = & \{-144, -174\}\coma  & \, \, \tilde{p}_i F^i     & = & 42\coma                     & N_{\rm flux} & = & 488\coma\\
   \langle g_s \rangle^{-1} & = & 190.44\coma     & \langle U^i \rangle & = & \{1.01298, 1.35064\} \coma  & |W_0|        & = & 1.04234 \cdot 10^{-4}\fstop\\
\end{array}  
\end{equation}
In addition, one thing worth noticing in these physical vacua of the non K3-fibered pCICYs is the significantly larger value of $N_{\rm flux}$, which may be hard to achieve in typical minimal global constructions. This could be used as an argument for filtering out the non K3-fibred CYs for the model building interests with the low $|W_0|$ values.


\section{Exponentially flat flux vacua (EFFV)}
\label{sec_EFFV}

We now discuss a special class of PFFV configurations which is protected against the non-perturbative (${\cal F}_{\rm inst}$) corrections due to an underlying symmetry of the CY $3$-fold, and we call them as exponentially flat flux vacua (EFFV). Given the correlation we have observed between the EFFV and the presence of the so-called ``non-trivially identical divisors" (NIDs)\footnote{The presence of NIDs facilitates an exchange symmetry in the model which is useful for constructing orientifolds with non-trivial odd-sector \cite{Gao:2013pra,Blumenhagen:2008zz, Carta:2020ohw, Altman:2021pyc, Gao:2021xbs, Carta:2022web} leading to interesting odd-axion phenomenology \cite{Lust:2006zg,Cicoli:2012vw,Gao:2013rra,Gao:2014uha,Carta:2021sms,Carta:2021uwv, Cicoli:2021tzt,Cicoli:2021gss}.} in our KS database analysis for CYs with $h^{1,1} = 2$ \cite{Carta:2021kpk}, it is a very natural to explore such possibilities in pCICY database. The very first reason is the fact that NIDs are very natural to get for most of the pCICYs as observed in \cite{Carta:2022web}, and therefore it is interesting to ask if this connection between the NIDs and EFFV has a deeper reason via investigating the underlying symmetry in the CICY geometry, and whether it can be extended to pCICYs with large $h^{1,1}$ as well.

\subsection{EFFV for pCICYs with $h^{1,1} = 2$}

These vacua fall under the category of the PFFV which we have studied in some good detail so far, which in addition satisfies: $M^1 = M^2$ or equivalently $H_1 + H_2 = 0$ for $h^{1,1} = 2$ case. One can directly read-off from the supersymmetric flatness condition in \cref{eq:SUSY-eff-M1=M2} that for exponential flat flux vacua, we simply need to demand the following conditions for the first order protection,
\begin{equation}
\label{eq:exp1}
H_1 = - H_2\coma n_1\, F^1 + n_2\, F^2 = 0\coma
\end{equation}
while one needs one additional condition in order to extend the protection against second order non-perturbative effects given as below,
\begin{equation}
\label{eq:exp2}
H_1 = - H_2\coma n_1\, F^1 + n_2\, F^2 = 0\coma F^1(2 {n_{11}}+{n_{12}})+{F^2} ({n_{12}}+2 {n_{22}}) = 0\fstop
\end{equation}
By considering the values of prepotential coefficients $n_i$ and $n_{ij}$ from Table \ref{tab:nijh112}, we find that all the $n_i$ and $n_{ij}$ coefficients are positive in addition to $n_i$'s being always integral valued while $n_{ij}$ can take rational values as well. Recall that using the orthogonality condition $M^i \, H_i = 0$, we have already fixed $F^1$ in terms of the remaining three fluxes, namely $\{F^2, H_1, H_2\}$ and one can check the consistency of those $F^1 \equiv F^1(F^2,H_1,H_2)$ expressions for each of the 36 pCICYs with \cref{eq:exp1,eq:exp2}. However, given that we have already collected the PFFV data, e.g., the one obtained from Step \ref{step1} in Table \ref{tab_fluxdata}, we can directly test \cref{eq:exp1,eq:exp2} on those PFFV configurations. In this investigation we find that although there are several CY examples with $n_1 \, F^1 = - n_2 \, F^2$, however, after demanding $M^1 = M^2$ as well, there remain only five CYs which have such PFFV configurations. Moreover, we find that those vacua also satisfy \cref{eq:exp2} ensuring the exponential protection of second order as well. The details are collected in Table \ref{tab:EFFV1}, which shows that there are a total of 9543 PFFV configurations which turns out to be exponentially flat. Also, because of the underlying symmetry, this protection can be anticipated to hold to all orders in the non-perturbative series of ${\cal F}_{\rm inst}$.

\begin{table}[H]
\centering
\begin{tabular}{|c||c|c|c|c|}
\hline
 CY \# & Total \# of PFFV & \cref{eq:exp1} & \cref{eq:exp2} & \cref{eq:exp1,eq:exp2} \\
 \hhline{|=#=|=|=|=}
 2 & 2913 & 2790 & 2790 & 2790 \\
 8 & 1348 & 1321 & 1321 & 1321 \\
 9 & 1348 & 1321 & 1321 & 1321 \\
 25 & 2913 & 2790 & 2790 & 2790 \\
 32 & 1905 & 1321 & 1321 & 1321 \\
\hline
\end{tabular}
\caption{Detailed list of total number of PFFV satisfying \cref{eq:exp1} or \cref{eq:exp2}.}
\label{tab:EFFV1}
\end{table}

\noindent
It may be worth to note that none of these 5 CYs are K3-fibred, and therefore these are not relevant for realizing low $|W_0|$ in the physically trustworthy regime, as we discussed in the previous section. 

\subsubsection*{Some more insights of the exponentially flat flux vacua}
Let us note that, there are only two fluxes (out of four) which remain independent for EFFV configurations because one flux $F^1$ for our convention is already determined by the orthogonality relation $M^i \, H_i = 0$ and one can eliminate, say, $H_1$ by $H_1 = - H_2$ leaving only $H_2$ and $F^2$ fluxes to remain independent. Moreover, given the fact that the flux rescaling redundancies in Step \ref{step2} demand $H_1$ and $H_2$ to be coprime, the condition $M^1 =M^2$ implies that such a reduced set of EFFV configurations have $H_1 = - 1 = - H_2$ or $H_1 = 1 = - H_2$ where we follow the former choice throughout this work. Therefore, there is only one flux parameter ($F^2$) which determines the distinct EFFV configurations.

The conditions discussed above altogether turn out to be quite strong as they significantly simplify the $N_{\rm flux}$ expression relevant for the D3 tadpole cancellation,
\begin{equation}
\label{eq:Nflux-M1eqM2}
N_{\rm flux} = - \frac{1}{2} F^i\, H_i = -\frac{1}{2} \left(F^1 H_1 + F^2 \, H_2 \right) = \frac{1}{2} (F^1 - F^2) > 0 \coma
\end{equation}
where in the last step we have used $H^1 = -1$ and $H^2 = 1$. Now as we have mentioned before, we have solved the orthogonality relation $M^i H_i = 0$ for $F^1$ which can be subsequently expressed in terms of three fluxes $H_1$, $H_2$ and $F^2$. So, this solution can be also used as a relation between $F^1$ and $F^2$ fluxes which holds for PFFV configurations in a given CY, for the fixed choice of $H_i$'s we have. It turns out that $F^1 = - m\, F^2$ where $m$ take the following rational values for all the 36 CICYs respectively,
\begin{equation}
\label{eq:F1=-mF2}
\begin{split}
m = & \,\left\{\frac{9}{5},1,\frac{9}{5},2,\frac{4}{3},\frac{14}{9},\frac{8}{7},1,1,3,\frac{16}{9},2,\frac{6}{5},3,\frac{7
   }{2},4,\frac{25}{16},3,4,\frac{19}{13},\frac{7}{2},\frac{9}{7},\frac{3}{2},\frac{17}{6},1,4,\frac{22}{13},\right.\\
& \,\left.4,\frac{10}{3},\frac{8}{3},\frac{15}{11},1,\frac{13}{4},4,\frac{5}{2},4\right\}\fstop
\end{split}
\end{equation}
\cref{eq:F1=-mF2} shows that there are five CYs for which $F^1 = - F^2$ (i.e., $m=1$) holds, and these are precisely the ones which have the symmetry in their respective CY geometries as listed in Table \ref{tab:EFFV1}, and are numbered as $\{2, 8, 9, 25, 32\}$ in the full list of 36 pCICYs. From \cref{eq:exp1,eq:exp2} and Table \ref{tab:nijh112}, one can read-off that $F^1 = - F^2$ is necessary for having the exponentially flat flux vacua, and for these EFFV one finds that $N_{\rm flux}$ can be simplified to be always given as $N_{\rm flux} = -F^2 > 0$ which follows from \cref{eq:Nflux-M1eqM2}. Also, we find that $M^i$ flux components, say $M^1 = M^2 = M$, for these five examples are respectively given as below
\begin{equation}
M = \left\{-\frac{1}{8\, F^2}, -\frac{1}{5\, F^2}, -\frac{1}{5\, F^2}, -\frac{1}{4\, F^2}, -\frac{1}{3\, F^2} \right\}\fstop
\end{equation}
In addition, we find that the quantities $\tilde{p}_{ij} F^j$ and $\tilde{p}_{i} F^i$ appearing in the integrality condition are given in terms of the $F^2$ flux as below,
\begin{equation}
\label{eq:integralityM1eqM2}
\begin{split}
    \tilde{p}_{ij} F^j & = \left(\left\{4 F^2, -4 F^2\right\}, \left\{\frac52 F^2, -\frac52 F^2\right\}, \left\{\frac52 F^2, -\frac52 F^2\right\}, \left\{2 F^2, -2 F^2\right\}, \left\{\frac32 F^2, -\frac32 F^2\right\} \right)\coma\\
\tilde{p}_{i} F^i &= \{0, 0, 0, 0, 0\}\fstop
\end{split}
\end{equation}
Thus, we find that for each value of $N_{\rm flux} = - F^2$, there are as many as EFFV as $F^2$ subject to satisfying the integrality condition in \cref{eq:integralityM1eqM2}, which means that CY \# 2 and 25 will have as many EFFV as one chooses the value of $F^2$, while for CY \# 8, 9 and 32 the number of EFFV is half of the choice of $F^2$ flux value.

\subsection{EFFV for pCICYs with $h^{1,1} \geq 3$}

For investigating the exponential flatness beyond the two-field case of $h^{1,1}=2$, we need to look at the general form of the dilaton dependent ``effective" superpotential $W^{\rm eff}(S)$ which is obtained by simplifying the generic flux superpotential along the perturbatively flat valley $U^i = S M^i$, and can be given as below \cite{Demirtas:2019sip, Carta:2021kpk},
\begin{equation}
\label{eq:W0eff}
W^{\rm eff}(S) = - \sqrt{\frac{2}{\pi}}\, {F}^{i} \, \partial_{i} {\cal F}_{\text{inst}}(S) \simeq - \sqrt{\frac{2}{\pi}}\, \sum_{d_i} \frac{A_{d_i}\, F^i\, d_i}{(2\pi\,i)^2}\, e^{2 \pi i \, S M^i d_i}\coma
\end{equation}
Using \cref{eq:W0eff}, one can anticipate that for having the exponential flatness along the dilaton direction one needs to impose the following condition,
\begin{equation}
\label{eq:Misequal}
M^i = M \in {\mathbb Q}^\ast \qquad \forall \, i\coma
\end{equation}
where ${\mathbb Q}^\ast$ denotes the set of non-zero rational numbers. In fact, if there exists a pair of components such that $M^i \neq M^j$ for some $i \neq j$, then one can perform the axio-dilaton stabilization similar to the $h^{1,1} =2$ case, irrespective of the fact whether it may (or may not) result in physical vacua and hence will accordingly lead to the corresponding PFFV being useful (or not) for phenomenological purposes. However, in any of the two such possibilities, the exponential flatness will be lifted. 

Let us also note that the simple looking condition in \cref{eq:Misequal} is quite restrictive in the sense that it demands identifications beyond merely finding the so-called NIDs at the level of topologies of the $4$-cycles inside the CY $3$-fold. Given that only 111 spaces out of 7820 favorable ones are a priory not suitable to have NIDs \cite{Carta:2022web}, this demand in  \cref{eq:Misequal} suggests that although the pCICY dataset is quite rich in having the NIDs, it would not be easy/frequent to find examples which can give EFFV. This will be better manifested after we present the complete analysis later on.

Now, while working with an appropriate basis and by considering the non-perturbative effects to the first order only, \cref{eq:W0eff} can be further expressed as below,
\begin{equation}
W^{\rm eff}(S) \simeq \frac{1}{(2 \pi)^2} \sqrt{\frac{2}{\pi}} \, \sum_{i} \, F^i \, n_i \, e^{2\pi i S \, M}\coma
\end{equation}
where the sum over $i$ runs in the $h^{1,1}$ of the mirror CYs we are using. Using this leading order piece along with the K\"ahler potential in \cref{eq:Keff}, the $F$-flatness condition in  \cref{eq:SUSY-eff} simplifies into the following form,
\begin{equation}
e^{- 2 \pi \,s \,  {M}} \,(1+ \pi\, s\, M) \,\sum_{i} \,  F^i\, n_i  = 0\fstop
\end{equation}
Therefore for the PFFV configurations which satisfy the requirement that all the components of the $M^i$ flux vector are identical, the flatness is protected at the first exponential order if the following condition holds,
\begin{equation}
M^i = M \quad \forall i\coma \sum_{i} F^i n_i = 0\fstop
\end{equation}
Recall that $M > 0$ follows from the K\"ahler cone conditions which imply that $(1+ \pi\, s\, M)$ cannot be set to zero for solving for the dilaton $s$. For the particular case of $h^{1,1} = 2$, we find  that $n_1 = n_2$ and $F^1 + F^2 = 0$ hold for the five pCICYs due to the underlying CY symmetry w.r.t. an exchange of the type $1 \leftrightarrow 2$ within various quantities such as divisor topologies (in terms of NIDs), intersection polynomial, the second Chern numbers and the GV invariants.

Now we check the exponential flatness against the second order effects so that to reduce the vacua with accidental flatness at the first order, which may (or may not) occur. For that purpose, we include the next order terms in the effective superpotential which takes the following form,
\begin{equation}
\begin{split}
W^{\rm eff}(S) \simeq &\,\frac{1}{(2 \pi)^2} \sqrt{\frac{2}{\pi}} \left[e^{2\pi i S \, M}\,\sum_{i} \, F^i \, n_i + e^{4\pi i S \, M} \sum_{\substack{i\\i \neq j}} F^i \, \left(2 n_{ii} + n_{ij}\right) \right]\fstop
\end{split}
\end{equation}
Now, the protection for the flatness will be extended to the second order if the following relations hold,
\begin{equation}
\label{eq:cond-EFFVs-2}
M^i = M \quad \forall i\coma \sum_{i} F^i \, n_i = 0\coma \sum_{\substack{i\\i \neq j}} F^i \, \left(2 n_{ii} + n_{ij}\right) = 0 \fstop
\end{equation}
Therefore if the $n_i$, $n_{ii}$ and $n_{ij}$ for $i \neq j$ are identical due to underlying symmetries of the CY $3$-fold, then it is sufficient to demand $\sum_{i} F^i = 0$ to guarantee EFFV configuration. 

We have investigated all the CY $3$-folds which have the exchange symmetry in terms of NIDs, and have further checked for those which have a unique topology so that to expect $n_i$ being equal (along with $M^i$'s) for all $i$. We find that there are only 31 such examples (out of 7820 favorable pCICYs), which means 26 more apart from the 5 examples of $h^{1,1}=2$ cases which we have analyzed in detail. We tabulate the values of $n_{i}$ and $n_{ij}$ for all these 31 CYs in Table \ref{tab:nijallequal}. We find that out of the 26 examples with $h^{1,1} \geq 3$, which have a single divisor topologies for each of them, there are 8 CYs which have $h^{1,1} =3$, 9 CYs have $h^{1,1} =4$, 6 CYs have $h^{1,1} =5$ while the remaining 3 CYs have $h^{1,1} =6$. Looking for EFFV, we find that although there are 8 pCICYs for $h^{1,1} = 3$ which have the possible exchange symmetry in terms of being NIDs but $n_i$ being equal holds only for three examples for which we have checked that $F^1 + F^2 +F^3 =0$ also holds. Given that a genuine exchange symmetry should respect the invariance of the SR ideal, intersection polynomial, $c_2(J_i)$ as well as within the GV invariants, we anticipate that merely having NIDs may not be enough for guaranteeing the exponential flat flux vacua. With this logic we conclude that there are only 13 pCICYs which can have EFFV of which 5 are those corresponding to $h^{1,1} =2$ while 8 corresponds to $h^{1,1} \geq 3$.

It has been argued in \cite{Blumenhagen:2009qh, Cicoli:2012fh} that there can be a superpotential contribution of the following kind arising either from gaugino condensation on spacetime-filling D3-branes or the Euclidean $D(-1)$-branes (also referred to as $E(-1)$-instantons), 
\begin{equation}
W_{\rm np}(S) \propto \, e^{i \, f\, S}\coma
\end{equation}
where $f$ is some flux dependent quantity, though it appears to be absent in generic models from the recent study of \cite{Kim:2022jvv}. However, note the fact that for the exponential flat flux vacua all the complex structure moduli are already aligned along the axio-dilaton direction which remains the only flat direction in the solution, and therefore the use of the fluxes in achieving the PFFV still have significant effects. The dilaton can receive VEVs from various contributions arising from several other sources such as string-loop corrections \cite{Berg:2005yu, Cicoli:2007xp} or $\alpha^\prime$-corrections \cite{Becker:2002nn} which can subsequently facilitate the complete minimization of the flat valley. 

Finally, let us mention that having less pCICYs resulting in EFFV configurations means finding the physical vacua through PFFV configurations is more likely and hence this observation about rarity of EFFV configurations should be considered in support of the overall goal of arguing the generality of PFFV vacua and KKLT scheme of moduli stabilization using pCICYs.


\section{Conclusions}
\label{sec:concl}

A dynamical realization of the low values for the GVW flux superpotential $|W_0|$ has been at the center of the type IIB orientifold models such as KKLT, and in this work, we have extended our previous study on flat vacua using the CY $3$-folds from the KS database \cite{Carta:2021kpk} to the mirror of the so-called pCICY $3$-folds. 

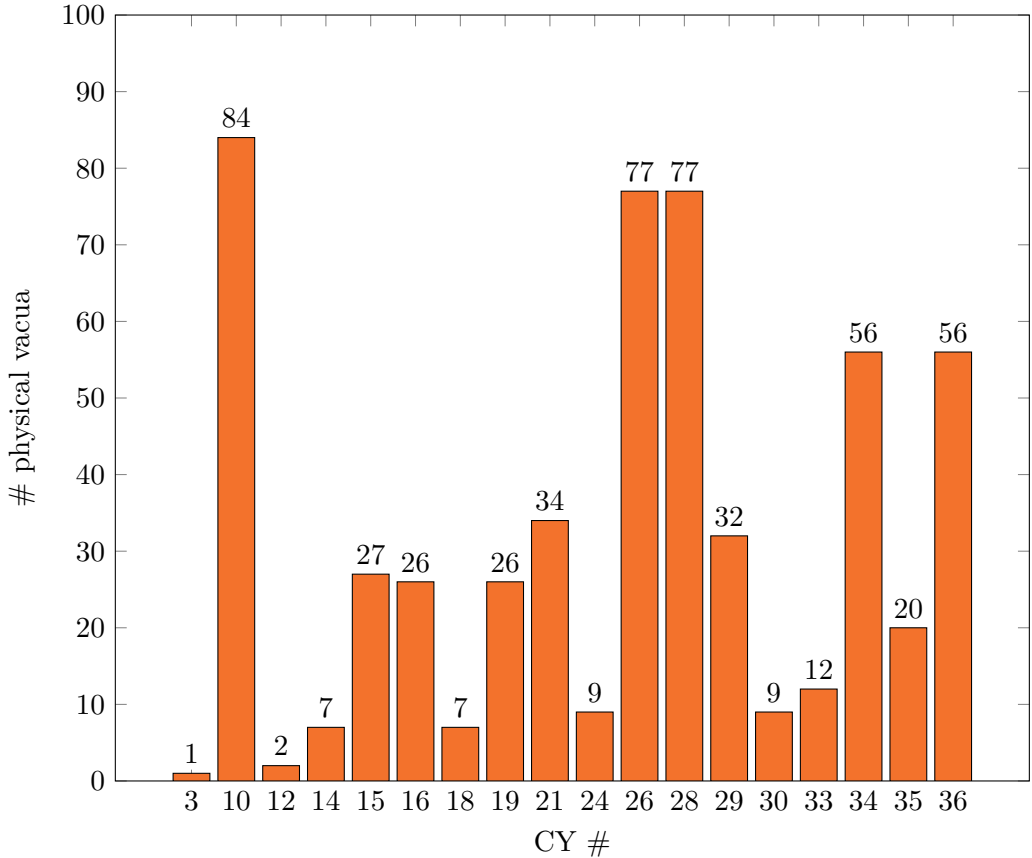
\begin{figure}[!htp]
 \centering
      \pgfplotstableread{
Label Nf  topper
 3  1  0 
 10  84  0 
 12  2  0 
 14  07  0 
 15  27  0 
 16  26  0 
 18  07  0 
 19  26  0 
 21  34  0 
 24  09  0 
 26  77  0 
 28  77  0 
 29  32  0 
 30  9  0 
 33  12  0 
 34  56  0 
 35  20  0 
 36  56  0 
    }\testdata
         \begin{tikzpicture}[scale=1]
    \begin{axis}[
       width=0.9\textwidth,
        ybar stacked,
        ymin=0,
        ymax=100,
        xtick={data},
        legend style={cells={anchor=west}, legend pos=north east},
        reverse legend=false, 
        xticklabels from table={\testdata}{Label},
        xticklabel style={text width=2cm,align=center},
      bar width=0.032\textwidth,
      xlabel={CY \#},
		ylabel={\# physical vacua},
    ]
    \addplot [fill=col2,ybar,nodes near coords,point meta=y]
            table [y=Nf, meta=Label, x expr=\coordindex]
                {\testdata};
    \end{axis}
    \end{tikzpicture}
\caption{PFFV configurations with physical vacua for all the 18 pCICY geometries.}
\label{fig:PFFVphysvacua}
\end{figure}

We first performed a systematic analysis of the PFFV using 36 pCICYs with $h^{1,1}=2$ for which the number of PFFV configurations obtained at various stages (and before fixing the dilaton valley) are given in \cref{tab_fluxdata}.  Subsequently, we classified the models into two categories, one which has mirror CY of a K3-fibered type and the other one which is not of the K3-fibered type. We observe that all the 16 K3-fibered CYs result in physical vacua while only 2 out of 20 CYs of non K3-fibered type give physical vacua. Thus, there are only 18 out of 36 pCICYs which give physical vacua via fixing the flat valley through non-perturbative effects ${\cal F}_{\rm inst}$. In the end, we find a total of 562 number of physical vacua arising from these 18 pCICYs, and the flux configurations corresponding to the lowest value of $|W_0|$ turns out to give $|W_0| \sim 10^{-27}$. The statistics of 562 physical vacua distributed to their respective pCICYs is presented in Figure \ref{fig:PFFVphysvacua}.

One of the main observations we make is the fact that the number of PFFV is dominated by the K3-fibered pCICYs, and in each of the four step, namely \ref{step1}-\ref{step4}, of our approach towards reaching the physical vacua, the percentage of contributions arising from the K3-fibered CY geometries increases. This is summarized in Table \ref{tab:fiberedpercen}. These results support the observations made from the KS database analysis in \cite{Carta:2021kpk}. 

\begin{table}[H]
\centering
\begin{tabular}{|c||c|c|c|}
\hline
Step & Total \# of PFFV & Total \# of PFFV from K3-fibered pCICYs & \% \\
\hhline{|=#=|=|=|}
\ref{step1} & 64520 & 48810 & 75.7 \\
\ref{step2} & 20095 & 17209 & 85.6 \\
\ref{step3} & 17662 & 16215 & 91.8 \\
\ref{step4} & 562 & 559 & 99.5 \\
\hline
\end{tabular}
\caption{Percentage of contributions arising from the K3-fibered CY geometries.}
\label{tab:fiberedpercen}
\end{table}

Moreover, given the fact that most of the pCICYs are K3-fibered and this number increases with increasing $h^{1,1}$ of the pCICY \cite{Anderson:2017aux}, one may anticipate to have a significant landscape of the perturbative flat flux vacua from the pCICY dataset which has a total of 7820 favorable CY $3$-folds. Although we do not have a clear analytic understanding about why the K3-fibered CYs result in more PFFV, and so far this is purely an empirical observation for us, however given that K3-fibration somehow reduces the number of non-zero triple intersection numbers by a theorem of Oguiso \cite{OGUISO:1993} (see \cite{Douglas:2003um} also) which translates into the fact that the K3 divisor can appear only linearly in the intersection polynomial, and hence many of the triple intersection numbers would vanish; for example see Table \ref{tab_data-h11=2} where the intersection numbers corresponding to the divisor topology T1$\equiv$K3 are listed. It can be seen that there are always two intersection numbers out of four which vanish. This can possibly result in a less restrictive set of conditions to be satisfied for realizing the flat vacua, e.g. via Step \ref{step1}. Therefore one can naively think that one of the reason for larger PFFVs for K3-fibered CYs should be rooted in this fact. However, a systematic analysis of pCICYs beyond the low $h^{1,1}$ is indeed desirable to be explored for seeing more possibilities or pattern given that for larger $h^{1,1}$, one may be able to make a further classification as whether the $K3$-fibered pCICYs are multiply-fibered and if there is a distinction in the PFFVs' pattern which one could observe. These are indeed interesting future directions one would like to pursue further.

We have also explored the possibilities of the so-called exponentially flat flux vacua (EFFV), and have found that there are five CICY $3$-folds with $h^{1,1} = 2$ which can have such vacua. We found that there is a total number of 9543 EFFV configurations for these five CY 3-folds, out of a total number of 64520 PFFV configurations arising from all the 36 pCICYs. We subsequently extended the search of finding such vacua for CICY geometries with larger $h^{1,1}$ which can have an underlying symmetry. To begin with, building on the learning from the $h^{1,1}=2$ cases, we anticipated that the presence of NIDs can be useful to find CYs which are suitable for having EFFV. However, the detailed analysis showed that there are additional restrictions arising from the identification of the components of the $M^i$ flux vector, which restricts the choice to only those CYs which have a single topology for all their coordinate divisors. In this context using the divisor topologies of pCICYs as presented in \cite{Carta:2022web} we find that apart from the five examples of $h^{1,1} = 2$, there are only 26 more pCICYs with $3 \leq h^{1,1} \leq 6$ with this property, and thus resulting in a total of 31 pCICYs (in the list of 7820 favorable geometries) which could be a priory suitable for giving EFFV. However, we further observed that despite having a single divisor topology in a given CY (i.e., having an exchange symmetry among all the divisors at the level of NIDs), the symmetry was not respected at the level of GV invariants dependent quantities $n_i$ and $n_{ij}$ defining the prepotential in \cref{eq:F-inst-nis}, as listed in Table \ref{tab:nijallequal}. The underlying reason for this mismatch at the GV invariant level can be correlated with the fact that intersection polynomials do not respect the symmetry. After taking this factor into account, we are left with only 8 pCICYs with $h^{1,1} \geq 3$, which can give exponentially flat flux vacua. However, let us emphasize that having less EFFVs means finding the physical vacua through PFFVs is more likely and hence supports the overall goal of arguing the generality of such vacua and KKLT scheme of moduli stabilization using pCICYs.

Finally, let us also mention that despite the extensive use in constructing MSSM-like models, the CY 3-folds of the pCICY dataset have been very rarely utilized for moduli stabilization and any subsequent phenomenological purposes such as realizing de-Sitter vacua and inflationary aspects, and our aim in this paper along with a companion paper \cite{Carta:2022web} has been to take some initiatives in these directions.


\section*{Acknowledgements}

F.C. is supported by STFC consolidated grant ST/T000708/1. A. M. is supported in part by Deutsche Forschungsgemeinschaft under Germany's Excellence Strategy EXC 2121 Quantum Universe 390833306. P. S. is thankful to Paolo Creminelli, Atish Dabholkar and Fernando Quevedo for their support.


\appendix

\section{pCICY geometries}
\label{App:pCICYgeom}

In this appendix, we briefly introduce the pCICYs and review their topologies. 

A pCICY is defined as the zero-locus of a set of $k$ homogeneous polynomials $p_j(z)$ in an ambient space $\mathcal{A}$ given by products of $\mathbb{P}^{n_i}$, satisfying
\begin{equation}
    \sum_{i}n_i-k=3\fstop
\end{equation}
They are usually identified by a configuration matrix that encodes the multi-degrees of the polynomial equations, i.e.,
\begin{equation}
\left[
\begin{tabular}{c|cccc}
$\PP^{n_1}$ &   $q_1^1$ & $\cdots$  & $q_k^1 $ \\
$\PP^{n_2}$  &   $q_1^2$ & $\cdots$  &$ q_k^2$  \\
$\vdots$ &   $\vdots$ & $\ddots$ & $\vdots$  \\ 
$\PP^{n_s} $&   $q_1^s $& $\cdots$ & $q_k^s $
\end{tabular}
\right] \fstop
\label{eq:configuration}
\end{equation}
The entries of the matrix are constrained by the condition on the vanishing condition for the first Chern class, so that the following equation holds
\begin{equation} \label{eq:CYpolynomialcondgen}
n_i+1=\sum_{j=1}^{k} q_j^i\coma \quad \forall \, i=1,\ldots,s \fstop
\end{equation}
A pCICY $X$ is said to be favorable if $h^{1,1}(X)=h^{1,1}(\mathcal{A})$. All pCICYs apart from 70 are favorable~\cite{Anderson:2017aux}. 

Using the configuration matrix in Eq. \eqref{eq:configuration}, it is possible to compute also the intersection polynomial of any favorable pCICY, lifting the integral of the triple intersection number $\kappa_{ijk}^0$ over the ambient space, i.e.
\begin{equation}
\kappa_{ijk}^0=\int_{\mathcal{A}} \hat{D}_i\wedge \hat{D}_j\wedge \hat{D}_k \bigwedge_{m=1}^k\left(\sum_{n=1}^s q_m^n \hat{D}_n\right)\coma
\label{eq:tripleintamb}
\end{equation}
where $\hat{D}_i$ is the Poincar\'e dual form of a divisor $D_i$ of $X$ descending from the hyperplane class of the corresponding $\PP^{n_i}$.

In \cite{Carta:2022web}, we classified all possible divisor topologies that appear in favorable pCICYs. The generic divisor topology for the favorable pCICYs turns out to be of the following form
\begin{equation}
\label{eq:divisor-topology}
D\equiv
\begin{tabular}{ccccc}
    & & 1 & & \\
   & 0 & & 0 & \\
$(\chi_{_h}-1)$ \quad & & $(\chi - 2 \chi_{_h})$ \quad & & \quad $(\chi_{_h} -1)$ \\
   & 0 & & 0 & \\
    & & 1 & & \\
\end{tabular}\coma
\end{equation}
where the generic expressions for $\chi(D)$ and $\chi_{_{h}}(D)$ are given as (e.g., see \cite{Blumenhagen:2008zz,Collinucci:2008sq, Bobkov:2010rf,Cicoli:2016xae})
\begin{equation}
\begin{split}
\chi(D) &= 2 h^{0,0} - 4 h^{1,0} + 2 h^{2,0} + h^{1,1}= \int_{X} \left(\hat{D} \wedge \hat{D} \wedge \hat{D} + c_2(X) \wedge \hat{D} \right)\coma \\
 \chi_{_h}(D) &= h^{0,0} - h^{1,0} + h^{2,0} = \frac{1}{12} \int_{X}\left(2\, \hat{D} \wedge \hat{D} \wedge \hat{D} + c_2(X) \wedge \hat{D} \right)\fstop \end{split}
\label{eq:chi-chih}
\end{equation}
We found that there are only 11 distinct topologies which arise from these pCICYs, that we classified in Table \ref{tab_divisor-topologies}.

\begin{table}[!htp]
\centering
\renewcommand{\arraystretch}{1.15}
\begin{tabular}{|Sc||Sc|Sc|Sc|Sc|Sc|} 
\hline
\shortstack{Sr.\\\#} & \shortstack{Divisor topology\\$\{h^{0,0}, h^{1,0}, h^{2,0}, h^{1,1}\}$} & \shortstack{frequency\\(57885 divisors)}  & \shortstack{frequency\\(7820 spaces)} & \shortstack{$h^{1,1}$ \\(pCICY)} & \makecell{\vspace{0.2cm} $\displaystyle{\int_{_{\rm CY}} \hat{D}^3}$} \\
\hhline{|=#=|=|=|=|=|} 
T1 & $K3 \equiv \{1, 0, 1, 20\}$ & 30901 & 7736 & 2-15 & 0 \\
T2 & $\{1, 0, 2, 30\}$ & 22150 & 7436 & 2-15 & 0 \\
T3 & $\{1, 0, 3, 38\}$ & 3372 & 2955 & 2-13 & 2 \\
T4 & $\{1, 0, 3, 36\}$ & 91 & 91 & 3-13 & 4 \\
T5 & $\{1, 0, 4, 46\}$ & 714 & 690 & 2-11 & 4 \\
T6 & $\{1, 0, 4, 45\}$ & 283 & 277 & 1-11 & 5 \\
T7 & $\{1, 0, 4, 44\}$ & 91 & 91 & 2-11 & 6 \\
T8 & $\{1, 0, 5, 52\}$ & 198 & 198 & 1-9 & 8 \\
T9 & $\{1, 0, 5, 51\}$ & 28 & 28 &  1-9 & 9 \\
T10 & $\{1, 0, 6, 58\}$ & 42 & 42 & 1-7 & 12 \\
T11 & $\{1, 0, 7, 64\}$ & 15 & 15 & 1-5 & 16 \\
\hline
\end{tabular}
\caption{Divisor topologies for favorable pCICYs and their frequencies of appearance, as we discussed in \cite{Carta:2022web}.}
\label{tab_divisor-topologies}
\end{table}

In the main text of this paper, we focused on constructing PFFV for compactification on the mirror pCICYs with $h^{2,1}=2$. For this reason, we collect all the topological data for favorable pCICYs with $h^{1,1} = 2$ in Table \ref{tab_data-h11=2}.

\begin{center}
\renewcommand{\arraystretch}{1.15}
  \begin{longtable}{|c||c|c||c|c|c|} 
 \caption{Topological data for favorable pCICYs with $h^{1,1} = 2$ taken from \cite{Carta:2022web}. The first number of pCICY in the second column is the corresponding number in the list of only favorable pCICY, while the second number is the number of the pCICY following \cite{Anderson:2017aux}. The classical triple intersection numbers $\kappa_{ijk}^0$ are collected as $\{\kappa_{111}^0, \kappa_{112}^0, \kappa_{122}^0, \kappa_{222}^0\}$, while other details for topologies Ti's are given in Table \ref{tab_divisor-topologies}.}  \\
\hline
\shortstack{${\cal M}_{i,j}$ \\$(i=h^{1,1})$} & \shortstack{Space \#\\$\{7820, 7890\}$}  & \makecell[c]{\vspace{0.3cm}$\{h^{1,1}, h^{2,1}\}$} & \shortstack{Topology\\of $J_i$} & \makecell[c]{\vspace{0.3cm}$\kappa_{ijk}^0$}  & \makecell[c]{\vspace{0.1cm}$\displaystyle{\int_{_{\rm CY}} c_2 \wedge J_i}$}\\
\hhline{|=#=|=#=|=|=|}
\endfirsthead
\hline
\shortstack{${\cal M}_{i,j}$ \\$(i=h^{1,1})$} & \shortstack{Space \#\\$\{7820, 7890\}$}  & \makecell[c]{\vspace{0.3cm}$\{h^{1,1}, h^{2,1}\}$} & \shortstack{Topology\\of $J_i$} & \makecell[c]{\vspace{0.3cm}$\kappa_{ijk}^0$}  & \makecell[c]{\vspace{0.1cm}$\displaystyle{\int_{_{\rm CY}} c_2 \wedge J_i}$}\\
\hhline{|=#=|=#=|=|=|}
\endhead
\label{tab_data-h11=2}${\cal M}_{2,1}$ & $\{7573, 7643\}$ & $\{2, 46\}$ & \{T2, T9\} & $\{0, 4, 12, 8\}$ & $\{36, 56\}$ \\
${\cal M}_{2,2}$ & $\{7574, 7644\}$ & $\{2, 46\}$ & \{T7, T7\} & $\{4, 12, 12, 4\}$ & $\{52, 52\}$ \\
${\cal M}_{2,3}$ & $\{7598, 7668\}$ & $\{2, 47\}$ & \{T2, T5\} & $\{0, 3, 9, 6\}$ & $\{36, 48\}$ \\
${\cal M}_{2,4}$ & $\{7655, 7725\}$ & $\{2, 50\}$ & \{T2, T10\} & $\{0, 4, 12, 12\}$ & $\{36, 60\}$ \\
${\cal M}_{2,5}$ & $\{7656, 7726\}$ & $\{2, 50\}$ & \{T4, T9\} & $\{2, 8, 12, 8\}$ & $\{44, 56\}$ \\
${\cal M}_{2,6}$ & $\{7688, 7758\}$ & $\{2, 52\}$ & \{T2, T7\} & $\{0, 4, 10, 4\}$ & $\{36, 52\}$ \\
${\cal M}_{2,7}$ & $\{7689, 7759\}$ & $\{2, 52\}$ & \{T4, T7\} & $\{2, 8, 10, 4\}$ & $\{44, 52\}$ \\
${\cal M}_{2,8}$ & $\{7691, 7761\}$ & $\{2, 52\}$ & \{T6, T6\} & $\{5, 10, 10, 5\}$ & $\{50, 50\}$ \\
${\cal M}_{2,9}$ & $\{7729, 7799\}$ & $\{2, 55\}$ & \{T4, T4\} & $\{2, 7, 7, 2\}$ & $\{44, 44\}$ \\
${\cal M}_{2,10}$ & $\{7736, 7806\}$ & $\{2, 56\}$ & \{T1, T5\} & $\{0, 0, 6, 6\}$ & $\{24, 48\}$ \\
${\cal M}_{2,11}$ & $\{7737, 7807\}$ & $\{2, 56\}$ & \{T2, T9\} & $\{0, 4, 10, 8\}$ & $\{36, 56\}$ \\
${\cal M}_{2,12}$ & $\{7738, 7808\}$ & $\{2, 56\}$ & \{T2, T8\} & $\{0, 3, 9, 9\}$ & $\{36, 54\}$ \\
${\cal M}_{2,13}$ & $\{7739, 7809\}$ & $\{2, 56\}$ & \{T4, T6\} & $\{2, 7, 9, 5\}$ & $\{44, 50\}$ \\
${\cal M}_{2,14}$ & $\{7746, 7816\}$ & $\{2, 58\}$ & \{T1, T9\} & $\{0, 0, 8, 8\}$ & $\{24, 56\}$ \\
${\cal M}_{2,15}$ & $\{7747, 7817\}$ & $\{2, 58\}$ & \{T1, T10\} & $\{0, 0, 8, 12\}$ & $\{24, 60\}$ \\
${\cal M}_{2,16}$ & $\{7749, 7819\}$ & $\{2, 58\}$ & \{T1, T11\} & $\{0, 0, 8, 16\}$ & $\{24, 64\}$ \\
${\cal M}_{2,17}$ & $\{7751, 7821\}$ & $\{2, 58\}$ & \{T2, T6\} & $\{0, 4, 8, 5\}$ & $\{36, 50\}$ \\
${\cal M}_{2,18}$ & $\{7752, 7822\}$ & $\{2, 58\}$ & \{T1, T9\} & $\{0, 0, 8, 8\}$ & $\{24, 56\}$ \\
${\cal M}_{2,19}$ & $\{7753, 7823\}$ & $\{2, 58\}$ & \{T1, T11\} & $\{0, 0, 8, 16\}$ & $\{24, 64\}$ \\
${\cal M}_{2,20}$ & $\{7763, 7833\}$ & $\{2, 59\}$ & \{T2, T4\} & $\{0, 3, 7, 2\}$ & $\{36, 44\}$ \\
${\cal M}_{2,21}$ & $\{7770, 7840\}$ & $\{2, 62\}$ & \{T1, T8\} & $\{0, 0, 6, 9\}$ & $\{24, 54\}$ \\
${\cal M}_{2,22}$ & $\{7774, 7844\}$ & $\{2, 62\}$ & \{T2, T4\} & $\{0, 4, 6, 2\}$ & $\{36, 44\}$ \\
${\cal M}_{2,23}$ & $\{7783, 7853\}$ & $\{2, 64\}$ & \{T2, T7\} & $\{0, 4, 8, 4\}$ & $\{36, 52\}$ \\
${\cal M}_{2,24}$ & $\{7788, 7858\}$ & $\{2, 66\}$ & \{T1, T6\} & $\{0, 0, 6, 5\}$ & $\{24, 50\}$ \\
${\cal M}_{2,25}$ & $\{7793, 7863\}$ & $\{2, 66\}$ & \{T4, T4\} & $\{2, 6, 6, 2\}$ & $\{44, 44\}$ \\
${\cal M}_{2,26}$ & $\{7797, 7867\}$ & $\{2, 68\}$ & \{T1, T10\} & $\{0, 0, 6, 12\}$ & $\{24, 60\}$ \\
${\cal M}_{2,27}$ & $\{7798, 7868\}$ & $\{2, 68\}$ & \{T2, T6\} & $\{0, 3, 7, 5\}$ & $\{36, 50\}$ \\
${\cal M}_{2,28}$ & $\{7799, 7869\}$ & $\{2, 68\}$ & \{T1, T10\} & $\{0, 0, 6, 12\}$ & $\{24, 60\}$ \\
${\cal M}_{2,29}$ & $\{7803, 7873\}$ & $\{2, 72\}$ & \{T1, T9\} & $\{0, 0, 6, 8\}$ & $\{24, 56\}$ \\
${\cal M}_{2,30}$ & $\{7812, 7882\}$ & $\{2, 76\}$ & \{T1, T7\} & $\{0, 0, 6, 4\}$ & $\{24, 52\}$ \\
${\cal M}_{2,31}$ & $\{7813, 7883\}$ & $\{2, 77\}$ & \{T2, T4\} & $\{0, 3, 5, 2\}$ & $\{36, 44\}$ \\
${\cal M}_{2,32}$ & $\{7814, 7884\}$ & $\{2, 83\}$ & \{T2, T2\} & $\{0, 3, 3, 0\}$ & $\{36, 36\}$ \\
${\cal M}_{2,33}$ & $\{7815, 7885\}$ & $\{2, 86\}$ & \{T1, T6\} & $\{0, 0, 4, 5\}$ & $\{24, 50\}$ \\
${\cal M}_{2,34}$ & $\{7816, 7886\}$ & $\{2, 86\}$ & \{T1, T9\} & $\{0, 0, 4, 8\}$ & $\{24, 56\}$ \\
${\cal M}_{2,35}$ & $\{7817, 7887\}$ & $\{2, 86\}$ & \{T1, T4\} & $\{0, 0, 4, 2\}$ & $\{24, 44\}$ \\
${\cal M}_{2,36}$ & $\{7818, 7888\}$ & $\{2, 86\}$ & \{T1, T9\} & $\{0, 0, 4, 8\}$ & $\{24, 56\}$ \\
\hline
\end{longtable}
\end{center}

\noindent
In constructing the complex structure prepotential for the mirror pCICYs we needed the computation of the Gopakumar-Vafa invariants of the pCICYs with $h^{1,1}=2$. They have been computed using \texttt{INSTANTON} \cite{Klemm:2001aaa}, with an algorithm reviewed in Appendix B of \cite{Carta:2021kpk} (see also \cite{Carta:2021sms}). In Table \ref{tab:nijh112} we list the coefficients entering the prepotential $\mathcal{F}_{\rm inst}$ in \cref{eq:F-inst-nis}.

\begin{center}
\renewcommand{\arraystretch}{1.15}
    \begin{longtable}{|c|c||c|c|c|c|c|c|}
    \caption{Coefficients of the prepotential $\mathcal{F}_{\rm inst}$ in \cref{eq:F-inst-nis} for pCICYs with $h^{1,1}=2$.} \\
    \hline 
    CY \# & pCICY \# & $n_1$ & $n_2$ & $n_{11}$ & $n_{12}$ & $n_{22}$ & Symmetry \\
    \hhline{|=|=#=|=|=|=|=|=|}
    \endfirsthead
    \hline 
    CY \# & pCICY \# & $n_1$ & $n_2$ & $n_{11}$ & $n_{12}$ & $n_{22}$ & Symmetry \\
    \hhline{|=|=#=|=|=|=|=|=|}
    \endhead
    \label{tab:nijh112}1 & $7643$ & $40$ & $144$ & $9$ & $496$ & $182$ & No \\
    2 & $7644$ & $64$ & $64$ & $28$ & $416$ & $28$ & Yes \\
    3 & $7668$ & $30$ & $225$ & $\frac{27}{4}$ & $747$ & $\frac{2025}{8}$ & No \\
    4 & $7725$ & $24$ & $144$ & $3$ & $432$ & $174$ & No \\
    5 & $7726$ & $40$ & $80$ & $5$ & $560$ & $30$ & No \\
    6 & $7758$ & $62$ & $140$ & $\frac{95}{4}$ & $780$ & $\frac{355}{2}$ & No \\
    7 & $7759$ & $70$ & $72$ & $\frac{67}{4}$ & $696$ & $35$ & No \\
    8 & $7761$ & $50$ & $50$ & $\frac{25}{4}$ & $650$ & $\frac{25}{4}$ & Yes \\
    9 & $7799$ & $80$ & $80$ & $25$ & $1127$ & $25$ & Yes \\
   10 & $7806$ & $24$ & $396$ & $3$ & $1152$ & $\frac{5319}{2}$ & No \\
   11 & $7807$ & $34$ & $140$ & $\frac{17}{4}$ & $692$ & $\frac{339}{2}$ & No \\
   12 & $7808$ & $18$ & $216$ & $\frac{9}{4}$ & $621$ & $243$ & No \\
   13 & $7809$ & $46$ & $84$ & $\frac{23}{4}$ & $865$ & $\frac{41}{2}$ & No \\
   14 & $7816$ & $32$ & $256$ & $4$ & $768$ & $1280$ & No \\
   15 & $7817$ & $16$ & $256$ & $2$ & $448$ & $1280$ & No  \\
   16 & $7819$ & $8$ & $256$ & $1$ & $256$ & $1280$ & No \\
   17 & $7821$ & $44$ & $144$ & $\frac{11}{2}$ & $1120$ & $158$ & No\\
   18 & $7822$ & $32$ & $256$ & $4$ & $768$ & $1280$ & No \\
   19 & $7823$ & $8$ & $256$ & $1$ & $256$ & $1280$ & No \\
   20 & $7833$ & $64$ & $213$ & $35$ & $1594$ & $\frac{1917}{8}$ & No \\
   21 & $7840$ & $12$ & $378$ & $\frac{3}{2}$ & $648$ & $\frac{10773}{4}$ & No \\
   22 & $7844$ & $80$ & $140$ & $18$ & $2016$ & $\frac{315}{2}$ & No \\
   23 & $7853$ & $64$ & $128$ & $10$ & $1216$ & $176$ & No \\
   24 & $7858$ & $36$ & $366$ & $\frac{9}{2}$ & $1584$ & $\frac{10863}{4}$ & No \\
   25 & $7863$ & $80$ & $80$ & $14$ & $1552$ & $14$ & Yes \\
   26 & $7867$ & $6$ & $360$ & $\frac{3}{4}$ & $360$ & $2727$ & No \\
   27 & $7868$ & $34$ & $204$ & $\frac{17}{4}$ & $1348$ & $\frac{459}{2}$ & No \\
   28 & $7869$ & $6$ & $360$ & $\frac{3}{4}$ & $360$ & $2727$ & No \\
   29 & $7873$ & $18$ & $348$ & $\frac{9}{4}$ & $900$ & $\frac{5499}{2}$ & No \\
   30 & $7882$ & $54$ & $336$ & $\frac{27}{4}$ & $2160$ & $2772$ & No \\
   31 & $7883$ & $72$ & $195$ & $10$ & $3195$ & $\frac{1755}{8}$ & No \\
   32 & $7884$ & $189$ & $189$ & $\frac{1701}{8}$ & $8262$ & $\frac{1701}{8}$ & Yes \\
   33 & $7885$ & $16$ & $640$ & $2$ & $2144$ & $10112$ & No \\
   34 & $7886$ & $4$ & $640$ & $\frac{1}{2}$ & $640$ & $10112$ & No \\
   35 & $7887$ & $64$ & $640$ & $8$ & $6912$ & $10112$ & No \\
   36 & $7888$ & $4$ & $640$ & $\frac{1}{2}$ & $640$ & $10112$ & No \\
 \hline
    \end{longtable}
\end{center}

\begin{center}
\renewcommand{\arraystretch}{1.15}
    \begin{longtable}{|c||c|c|c|c|}
    \caption{Details of ${\cal F}_{\rm inst}$ (defined in \cref{eq:F-inst-nis}) for all the 31 CYs which are relevant to examine the EFFV are presented, along with the presence of an underlying symmetry. The counting of $n_i$ also determines the $h^{1,1}$ for the corresponding CY. It is worth noticing that the first example has all the $n_i$'s being equal, while the symmetry does not hold for higher order GV dependent quantities, $n_{ii}$ and $n_{ij}$.}\\
    \hline 
     \# & $n_i$ & $n_{ii}$ & $n_{i<j}$ & Symmetry \\
    \hhline{|=#=|=|=|=|}
    \endfirsthead
    \hline 
     \# & $n_i$ & $n_{ii}$ & $n_{i<j}$ & Symmetry \\
    \hhline{|=#=|=|=|=|}
    \endhead
 \label{tab:nijallequal}$3412 $& $\{8,8,8,8,8,8\}       $& $\{1,1,1,1,1,1\}                                                                        $& $\{6,6,8,0,0,6,0,8,0,0,0,8,6,6,6\} $& no  \\
  $3413 $& $\{9,9,9,9,9,9\}       $& $\left\{\frac{9}{8},\frac{9}{8},\frac{9}{8},\frac{9}{8},\frac{9}{8},\frac{9}{8}\right\} $& $\{6,6,6,0,0,0,0,6,6,0,6,6,6,6,0\} $& yes \\
  $4111 $& $\{18,20,8,8,8\}       $& $\left\{\frac{9}{4},\frac{5}{2},1,1,1\right\}                                           $& $\{20,24,0,0,0,12,12,6,6,6\}       $& no  \\
  $5299 $& $\{54,54,54\}          $& $\left\{\frac{27}{4},\frac{27}{4},\frac{27}{4}\right\}                                  $& $\{54,54,54\}                      $& yes \\
  $5302 $& $\{16,16,16,16,16,16\} $& $\{2,2,2,2,2,2\}                                                                        $& $\{8,8,8,8,8,8,8,8,8,8,8,8,8,8,8\} $& yes \\
  $5304 $& $\{18,36,9,18\}        $& $\left\{\frac{9}{4},\frac{9}{2},\frac{9}{8},\frac{9}{4}\right\}                         $& $\{36,18,0,12,36,18\}              $& no  \\
  $5305 $& $\{18,18,18,18\}       $& $\left\{\frac{9}{4},\frac{9}{4},\frac{9}{4},\frac{9}{4}\right\}                         $& $\{18,36,0,21,36,18\}              $& no  \\
  $5829 $& $\{14,14,16,8,8\}      $& $\left\{\frac{7}{4},\frac{7}{4},2,1,1\right\}                                           $& $\{22,4,12,0,4,0,12,10,10,6\}      $& no  \\
  $6024 $& $\{15,15,15,9,9\}      $& $\left\{\frac{15}{8},\frac{15}{8},\frac{15}{8},\frac{9}{8},\frac{9}{8}\right\}          $& $\{3,3,9,9,3,9,9,9,9,0\}           $& no  \\
  $6222 $& $\{42,20,14,14\}       $& $\left\{\frac{45}{4},\frac{5}{2},\frac{7}{4},\frac{7}{4}\right\}                        $& $\{20,12,12,19,19,22\}             $& no  \\
  $6223 $& $\{18,32,14,14\}       $& $\left\{\frac{9}{4},4,\frac{7}{4},\frac{7}{4}\right\}                                   $& $\{32,36,0,7,40,22\}               $& no  \\
  $6770 $& $\{32,32,32,12,12\}    $& $\left\{4,4,4,\frac{3}{2},\frac{3}{2}\right\}                                           $& $\{16,16,32,32,16,32,32,32,32,0\}  $& no  \\
  $6971 $& $\{36,48,36\}          $& $\left\{\frac{27}{2},6,\frac{27}{2}\right\}                                             $& $\{144,36,144\}                    $& no  \\
  $7073 $& $\{26,26,14,14\}       $& $\left\{\frac{13}{4},\frac{13}{4},\frac{7}{4},\frac{7}{4}\right\}                       $& $\{28,34,7,7,34,19\}               $& no  \\
  $7242 $& $\{28,28,28\}          $& $\left\{\frac{7}{2},\frac{7}{2},\frac{7}{2}\right\}                                     $& $\{48,48,48\}                      $& yes \\
  $7247 $& $\{24,24,24,9\}        $& $\left\{3,3,3,\frac{9}{8}\right\}                                                       $& $\{30,30,12,30,12,12\}             $& no  \\
  $7447 $& $\{24,24,24,24,24\}    $& $\{3,3,3,3,3\}                                                                          $& $\{24,24,24,24,24,24,24,24,24,24\} $& yes \\
  $7487 $& $\{24,24,24,24,24\}    $& $\{3,3,3,3,3\}                                                                          $& $\{24,24,24,24,24,24,24,24,24,24\} $& yes \\
  $7489 $& $\{36,20,20,8\}        $& $\left\{\frac{9}{2},\frac{5}{2},\frac{5}{2},1\right\}                                   $& $\{20,20,24,34,13,13\}             $& no  \\
  $7580 $& $\{72,36,18\}          $& $\left\{9,\frac{9}{2},\frac{9}{4}\right\}                                               $& $\{240,72,72\}                     $& no  \\
  $7581 $& $\{36,39,54\}          $& $\left\{\frac{9}{2},\frac{39}{8},\frac{99}{4}\right\}                                   $& $\{153,54,117\}                    $& no  \\
  $7590 $& $\{18,18,18,18\}       $& $\left\{\frac{9}{4},\frac{9}{4},\frac{9}{4},\frac{9}{4}\right\}                         $& $\{36,36,3,3,36,36\}               $& no  \\
  $7644 $& $\{64,64\}             $& $\{28,28\}                                                                              $& $\{416\}                           $& yes \\
  $7669 $& $\{36,36,36\}          $& $\left\{\frac{9}{2},\frac{9}{2},\frac{9}{2}\right\}                                     $& $\{117,117,117\}                   $& yes \\
  $7729 $& $\{60,33,33\}          $& $\left\{\frac{27}{2},\frac{33}{8},\frac{33}{8}\right\}                                  $& $\{123,123,114\}                   $& no  \\
  $7761 $& $\{50,50\}             $& $\left\{\frac{25}{4},\frac{25}{4}\right\}                                               $& $\{650\}                           $& yes \\
  $7799 $& $\{80,80\}             $& $\{25,25\}                                                                              $& $\{1127\}                          $& yes \\
  $7846 $& $\{54,54,21\}          $& $\left\{\frac{27}{4},\frac{27}{4},\frac{21}{8}\right\}                                  $& $\{204,135,135\}                   $& no  \\
  $7862 $& $\{48,48,48,48\}       $& $\{6,6,6,6\}                                                                            $& $\{160,160,160,160,160,160\}       $& yes \\
  $7863 $& $\{80,80\}             $& $\{14,14\}                                                                              $& $\{1552\}                          $& yes \\
  $7884 $& $\{189,189\}           $& $\left\{\frac{1701}{8},\frac{1701}{8}\right\}                                           $& $\{8262\}                          $& yes \\
 \hline
    \end{longtable}
\end{center}


\bibliographystyle{JHEP}
\bibliography{reference}

\end{document}